\documentclass{cernrep}

\newcommand{\beq}{\begin{equation}}
\newcommand{\eeq}{\end{equation}}
\newcommand{\bea}{\begin{eqnarray}}
\newcommand{\eea}{\end{eqnarray}}
\newcommand{\bear}{\begin{eqnarray}}
\newcommand{\eear}{\end{eqnarray}}

\newcommand{\ba}{\begin{array}}
\newcommand{\ea}{\end{array}}

\newcommand{\bpm}{\begin{pmatrix}}
\newcommand{\epm}{\end{pmatrix}}

\begin{document}
\title{Three Lectures of Flavor and CP violation within and Beyond the Standard Model}
\author{ S. Gori}
\institute{Department of Physics, University of Cincinnati, Cincinnati, Ohio, USA}
\maketitle

\begin{abstract}
In these lectures I discuss 1) flavor physics within the Standard Model, 2) effective field theories and Minimal Flavor Violation, 3) flavor physics in theories beyond the Standard Model and ``high energy" flavor transitions of the top quark and of the Higgs boson. As a bi-product, I present the most updated constraints from the measurements of $B_s\to\mu^+\mu^-$, as well as the most recent development in the LHC searches for top flavor changing couplings. 
\end{abstract}


\section{Introduction}\label{sec:intro}

My plan for these lectures is to introduce you to the basics of flavor physics and CP violation. These three lectures that I gave at the 2015 European School of High-Energy Physics are not comprehensive, but should serve to give an overview of the interesting open questions in flavor physics and of the huge experimental program measuring flavor and CP violating transitions. Hopefully they will spark your curiosity to learn more about flavor physics. There are many books and reviews about flavor physics for those of you interested \cite{Buras:1998raa,Brancobook,Manohar:2000dt,Nir:2005js,Grinstein:2015nya,Ligeti:2015kwa}.

Flavor physics is the study of different generations, or ``flavors", of quarks and leptons, their spectrum and their transitions. There are six different types of quarks: up ($u$), down ($d$), strange ($s$), charm ($c$), bottom ($b$) and top ($t$) and three different type of charged leptons: electron, muon and tau. In these lectures, I will concentrate on the discussion of quarks and the mesons that contain them. A recent review about lepton flavor violation can be found in \cite{deGouvea:2013zba}.

The Large Hadron Collider (LHC) discovery of the Higgs boson in 2012 \cite{Aad:2012tfa,Chatrchyan:2012xdj} and the subsequent early measurements of its couplings to the Standard Model (SM) gauge bosons and third generation quarks and leptons have been a remarkably successful confirmation of the SM and of its mechanism of electroweak symmetry breaking (EWSB). The LHC has been able to demonstrate that the Higgs does not couple universally with (some) quarks and leptons already with Run I data \cite{Khachatryan:2016vau}. In fact, we know that in the SM $m_t\gg m_c\gg m_u$ and $m_b\gg m_s\gg m_d$, and that the same hierarchies hold for the Higgs Yukawa couplings with quarks and leptons. Our lack of understanding of why nature has exactly three generations of quarks and leptons and why their properties (masses and mixing angles) are described by such hierarchical values is the so called ``Standard Model flavor puzzle".
In the limit of unbroken electroweak (EW) symmetry none of the basic constituent of matter would have a non-zero mass. The SM flavor puzzle is, therefore, intimately related to the other big open question in particle physics, i.e. which is the exact mechanism behind EWSB.

Once the SM quark and lepton masses, as well as quark mixing angles (3 plus a phase) have been fixed, the SM is a highly predictive theory for flavor transitions. Particularly, any flavor transition has to involve the exchange of at least a $W$ boson and therefore flavor changing neutral transitions can only arise (at least) at the loop-level.  In the last few years, tremendous progress has been reached in testing the mechanism of quark flavor mixing by several experiments (LHCb and B-factories (Belle and Babar) as well as the high energy experiments ATLAS and CMS), finding good agreement with the SM expectations.
At the same time, there are a few flavor measurements that could be interpreted as tantalizing hints for deviations if compared to the SM predictions.
Particularly, lately there have been a a lot of attention on the anomalies in angular observables in the decay $B_d\to K^*\mu\mu$ (involving a $b\to s$ flavor transition), as observed by the LHCb collaboration \cite{Aaij:2013qta,Aaij:2015oid,Abdesselam:2016llu}, as well as on the observables testing lepton flavor universality, BR$(B\to K\mu\mu)/{\rm{BR}}(B\to K ee)$, as observed at LHCb \cite{Aaij:2014ora} and on the rare decays $B\to D\tau\nu_\tau$ and $B\to D^* \tau\nu_\tau$ by Belle, Babar, and LHCb \cite{Lees:2012xj,Lees:2013uzd,Aaij:2015yra,Huschle:2015rga,Abdesselam:2016cgx}.

The coming years will be exciting since several low (and high) energy flavor experiments will collect a lot more data. In particular \cite{Ligeti:2015kwa}, 
\beq
\frac{{\rm{LHCb~upgrade}}}{{\rm{LHCb}},~1{\rm{fb}}^{-1}}\sim \frac{{\rm{Belle~ II~data}}}{{\rm{Belle~data}}}\sim 50,~~~~\frac{{\rm{HL-LHC}}}{{\rm{LHC,~ICHEP~2016}}}\sim 200
\eeq
in the time scale of $\sim 20$ years for the LHCb upgrade and for the High-Luminosity LHC and of $\sim 10$ years for Belle II.

Present and future flavor measurements will be able to probe, and eventually indirectly discover, New Physics (NP). Observing new sources of flavor mixing is, in fact, a natural expectation for any extension of the SM with new degrees of freedom not far from the TeV scale. While direct searches of new particles at high energies provide information on the mass spectrum of the possible new degrees of freedom, the indirect information from low energy flavor observables translates into unique constraints on their couplings.

The lectures are organized as follows: In Sec. \ref{Sec:SMflavor}, I will introduce
the main ingredients of flavor physics and CP violation in the SM. I will both review the theory aspects and the experimental determination of the several SM flavor parameters. My second lecture, in Sec. \ref{Sec2}, will discuss the role of flavor physics in testing effective field theories beyond the SM (BSM), where new degrees of freedom are heavy if compared to the EW scale, and they can be integrated out, to generate higher dimensional operators to be added to the SM Lagrangian. Sec. \ref{Sec3} is dedicated to the discussion of the flavor properties of specific BSM theories, i.e. models with multi-Higgs doublets and Supersymmetric models. I will also discuss the interplay between low energy flavor measurements and high energy flavor measurements involving top and Higgs flavor transitions, as it can be measured at ATLAS and CMS. Finally, I will conclude in Sec. \ref{sec:conclusions}.

\section{Flavor physics in the Standard Model}\label{Sec:SMflavor}

\subsection{The flavor sector of the Standard Model}\label{sec:SMFlavor}

The Standard Model (SM) Lagrangian can be divided in three main parts: the gauge, the Higgs, and the flavor sector. The first two parts are highly symmetric

\bea\nonumber
\mathcal L _{\rm{SM}}^{\rm{gauge}}+\mathcal L _{\rm{SM}}^{\rm{Higgs}}&=&i\sum_i\sum_\psi \bar\psi^i \slashed{D}\psi^i-\frac{1}{4}\sum_a G_{\mu\nu}^aG_{\mu\nu}^a-\frac{1}{4}\sum_a W_{\mu\nu}^aW_{\mu\nu}^a+\\\label{eq:LgaugeH}
&&-\frac{1}{4} B_{\mu\nu}B_{\mu\nu}+|D_\mu \phi|^2+(\mu^2|\phi|^2-\lambda|\phi|^4),
\eea
and fully determined by a small set of free parameters: the three gauge couplings, $g_3,g_2,g_1$ corresponding to the SM gauge groups $SU(3)\times SU(2)\times U(1)_Y$, the Higgs ($\phi$) mass, $m_h$, and the Higgs vacuum expectation value (VEV), $v$ (or, equivalently, the Higgs mass term, $\mu$, and the quartic coupling, $\lambda$). In this expression $G,W$, and $B$ are the SM $SU(3)$, $SU(2)$, and $U(1)_Y$ gauge fields, respectively, and we have defined the quark and lepton field content, $\psi^i$, as 
\bea\label{eq:matterContent}
&&\psi^i\equiv Q_L^i, L_L^i, u_R^i, d_R^i, e_R^i,~~~~{\rm{with}}\\\nonumber
&& Q_L^i=(3,2,1/6),~L_L^i=(1,2,-1/2),~u_R^i=(3,1,2/3),~d_R^i=(3,1,-1/3),~e_R^i=(1,1,-1),
\eea
where $i=1,2,3$ is the flavor (or generation) index and the three numbers refer to the representation under the SM gauge group. 

The Lagrangian in (\ref{eq:LgaugeH}) possesses a large flavor symmetry that can be decomposed as 
\beq\label{eq:flavorSymm}
\mathcal G_{\rm{flavor}}=SU(3)^5\times U(1)^5=SU(3)^3_q\times SU(3)^2_\ell\times U(1)_B\times U(1)_L\times U(1)_Y\times U(1)_{\rm{PQ}}\times U(1)_E,
\eeq
where three $U(1)$ symmetries can be identified with baryon and lepton numbers, and hypercharge, the latter of which is broken spontaneously by the Higgs field. The two remaining $U(1)$ groups can be identified with the Peccei-Quinn symmetry \cite{Peccei:1977ur} and with a global rotation of a single $SU(2)$ singlet ($e_R$ in the case of Eq. (\ref{eq:flavorSymm})). The flavor sector of the SM Lagrangian breaks the $SU(3)^5$ symmetry through the Yukawa interactions
\beq\label{eq:Yukawas}
\mathcal L_{\rm{yuk}}=-Y_d^{ij}\bar Q_L^i\phi D_R^j-Y_u^{ij}\bar Q_L^i\tilde \phi U_R^j-Y_e^{ij}\bar L_L^i\phi e_R^j+{\rm{h.c.}},
\eeq
where $\phi$ is the Higgs field ($\phi=(1,2,1/2)$), $\tilde\phi$ is its conjugate representation $\tilde\phi=i\tau_2\phi^\dagger$ and $Y_{d,u,e}$ are the three Yukawa couplings. 

The diagonalization of each Yukawa coupling requires a bi-unitary transformation. Particularly, in the absence of right-handed (RH) neutrinos as in Eq. (\ref{eq:matterContent}), the lepton sector Yukawa can be fully diagonalized by the transformation $U_{eL}Y_eU_{eR}^\dagger={\rm{diag}}(y_e^1,y_e^2,y_e^3)=\sqrt 2 ~{\rm{diag}}(m_e,m_\mu,m_\tau)/v$. In the quark sector, it is not possible to simultaneously diagonalize the two Yukawa matrices $Y_u$ and $Y_d$ without breaking the $SU(2)$ gauge invariance. If, for example, we choose the basis in which the up Yukawa is diagonal, then 
\beq
Y_u={\rm{diag}}(y_u^1,y_u^2,y_u^3)=\frac{\sqrt 2}{v}(m_u,m_c,m_t),~~Y_d=V\cdot{\rm{diag}}(y_d^1,y_d^2,y_d^3)=\frac{\sqrt 2}{v}V\cdot(m_d,m_s,m_b),
\eeq
where we have defined the Cabibbo-Kobayashi-Maskawa (CKM) matrix as $V=U_{uL}U_{dL}^\dagger$.

However, in the SM the $SU(2)$ gauge symmetry is broken spontaneously by the Higgs field and therefore, we can equivalently rotate both left-handed (LH) up and down quarks independently, diagonalizing simultaneously up and down quark masses. By performing these transformations, the CKM dependence moves into the couplings of up and down quarks with the $W$ boson. In particular, the charged-current part of the quark covariant derivative in (\ref{eq:LgaugeH}) can be rewritten in the mass eigenstate basis as
\beq
-\frac{g}{2}\bar Q_L^i\gamma^\mu W_\mu^a\tau^a Q_L^i\xrightarrow{{\rm{mass-basis}}}-\frac{g}{\sqrt 2}(
\begin{array}{ccc}
\bar u_L & \bar c_L & \bar t_L\end{array})\gamma^\mu W_\mu^+ V \left(
\begin{array}{c}
d_L\\
s_L\\
b_L
\end{array}\right).
\eeq
This equation shows that the appearance of $W$ boson flavor changing couplings. This is the only flavor changing interaction in the SM. {\underline{{\bf{Exercise:}}}} prove that the neutral interactions of the photon, the $Z$ boson, the gluons and the Higgs boson are flavor diagonal in the quark mass eigenbasis. We can therefore conclude that, in the SM, 

\begin{itemize}
\item[(a)] the only interactions mediating flavor changing transitions are the charged interactions;
\item[(b)] there are no tree-level flavor changing neutral interactions.
\end{itemize}

In spite of point (a), it must be stressed that $V$, the CKM matrix, originates from the Yukawa sector: in absence of Yukawa couplings, $V_{ij} =\delta_{ij}$ and therefore we have no flavor changing transitions.

We can now count the number of free parameters of the SM Lagrangian. As opposed to the five free parameters of the gauge and Higgs sector ($g_1,g_2,g_3,v,m_h$), the flavor part of the Lagrangian has a much larger number of free parameters. Particularly, the CKM matrix is defined by 4 free parameters: three real angles and one complex CP-violating phase. {\underline{\bf{Exercise:}} Using the unitarity relations discussed in the next subsection, demonstrate that the CKM matrix is fully described by 4 free parameters. This phase is the only source of CP violation in the SM, beyond the QCD phase, $\theta_{\rm{QCD}}$.
The full set of parameters controlling the breaking of the quark flavor symmetry is composed by six quark masses and four parameters of CKM matrix (to be added to the three charged lepton masses, as obtained from the Yukawa coupling $Y_e$). 

Many parameterizations of the CKM matrix have been proposed in the literature. In these lectures, we will focus on the standard parametrization \cite{Chau:1984fp} and on the Wolfenstein parametrization \cite{Wolfenstein:1983yz}.   
The CKM matrix is unitary and can be described by three rotation angles $\theta_{12},\theta_{13},\theta_{12}$ and a complex phase $\delta$. In all generality, we can write the standard parametrization as product of three rotations with respect to three orthogonal axes
\begin{eqnarray}\nonumber
V&=&\left(\begin{array}{ccc}
V_{ud} & V_{us} &  V_{ub}\\
V_{cd} & V_{cs} &  V_{cb}\\
V_{td} & V_{ts} &  V_{tb}\end{array}\right)=
\left(\begin{array}{ccc}
1 & 0 &0\\
0 & c_{23} & s_{23}\\
0 & -s_{23} & c_{23}\end{array}\right)
\left(\begin{array}{ccc}
c_{13} & 0 & s_{13}e^{-i\delta}\\
0 & 1 & 0\\
-s_{13}e^{i\delta} & 0 & c_{13}\end{array}\right)
\left(\begin{array}{ccc}
c_{12} & s_{12} & 0\\
-s_{12} & c_{12} & 0 \\
0 & 0 & 1\end{array}\right)
=\\
&=&\left(\begin{array}{ccc}
c_{12}c_{13}&s_{12}c_{13}&s_{13}e^{-i\delta}\\ -s_{12}c_{23}
-c_{12}s_{23}s_{13}e^{i\delta}&c_{12}c_{23}-s_{12}s_{23}s_{13}e^{i\delta}&
s_{23}c_{13}\\ s_{12}s_{23}-c_{12}c_{23}s_{13}e^{i\delta}&-s_{23}c_{12}
-s_{12}c_{23}s_{13}e^{i\delta}&c_{23}c_{13}
\end{array}\right),
\end{eqnarray}
where we have denoted $s_{ij}\equiv \sin\theta_{ij}$ and $c_{ij}\equiv \cos\theta_{ij}$, $i,j=1,2,3$.

From measurements, we know that $s_{12}$, $s_{13}$ and $s_{23}$ are small numbers, therefore we can approximately write the CKM matrix in terms of an expansion in $|V_{us}|$
\begin{equation}\label{CKMWolf} 
V=
\left(\begin{array}{ccc}
1-{\lambda^2\over 2}&\lambda&A\lambda^3(\varrho-i\eta)\\ -\lambda&
1-{\lambda^2\over 2}&A\lambda^2\\ A\lambda^3(1-\varrho-i\eta)&-A\lambda^2&
1\end{array}\right)
+\mathcal O(\lambda^4)\,,
\end{equation}
with $\lambda\sim 0.23$ the Cabibbo angle and the parameters $A,\rho,\eta$ of the order 1, defined as
\beq
\lambda\equiv s_{12},~A\lambda^2\equiv s_{23},~A\lambda^3(\rho-i\eta)\equiv s_{13}e^{-i\delta}.\eeq
This is the Wolfenstein parametrization, that shows clearly the sizable hierarchies in between the several elements of the CKM matrix, that, at the zeroth order in $\lambda$ is given by the identity matrix.

\subsection{Tests of the CKM matrix}

The unitarity of the CKM matrix implies the following relations between its elements:
\beq\label{eq:unitarityRelations}
{\rm{Phase~independent:}} \sum_{k=1,2,3}|V_{ik}|^2=1, ~~~{\rm{Phase~dependent:}} \sum_{k=1,2,3}V_{ki}V_{kj}^*=0,~~ j\neq i.
\eeq

These relations are a distinctive feature of the SM, where the CKM matrix is the only source of quark flavor transitions. Each of the phase dependent relations, for fixed $i$ and $j$, can be
visualized as a triangle in the complex plane, where each side represents the complex number $V_{ki}V_{kj}^*$ for the three different $k = u, c,t$. The fact that the three vectors add up
to form a closed triangle is the manifestation of the unitarity relation.
Among the six phase dependent relations, the most stringent test is provided by the $i=1$ and $j=3$ case, since, in this case, the corresponding unitarity triangle has all sides of the same order in $\lambda$. Particularly, the unitarity relation can be written as
\beq
\frac{V_{ud}V_{ub}^*}{V_{cd}V_{cb}^*}+\frac{V_{td}V_{tb}^*}{V_{cd}V_{cb}^*}+1=0 ~~~~\leftrightarrow ~~~~(\bar\rho+i\bar\eta)+(1-\bar\rho-i\bar\eta)+1=0,
\eeq
where this defines the parameters $\bar\rho$ and $\bar\eta$, which are approximately given by
\beq
\bar\rho\simeq\rho\left(1-\frac{\lambda^2}{2}\right),~~\bar\eta\simeq \eta\left(1-\frac{\lambda^2}{2}\right).
\eeq
This unitarity triangle is represented on the left panel of Fig.~\ref{fig:CKM}. We have defined the angles of this triangle
\bea\label{angledef}
\alpha &\equiv& \arg\left(-{V_{td}V_{tb}^*\over V_{ud}V_{ub}^*}\right), \quad
  \beta \equiv \arg\left(-{V_{cd}V_{cb}^*\over V_{td}V_{tb}^*}\right), \quad
  \gamma \equiv \arg\left(-{V_{ud}V_{ub}^*\over V_{cd}V_{cb}^*}\right),
\eea
and also we can define one additional angle $\beta_s$ as
\beq
\beta_s \equiv \arg\left( -\frac{V_{ts}V_{tb}^*}{V_{cs}V_{cb}^*}\right).
\eeq

There are many measurements performed at different experiments (Babar, Belle, LHCb) that over-constrain the values of the elements of the CKM matrix. In the right panel of Fig.~\ref{fig:CKM}, we report a summary of the most stringent experimental constraints on the several CKM elements. Every element but $V_{td}$ and $V_{ts}$ are determined directly by tree-level processes. In particular

\begin{itemize}
\item $V_{ud}$ is extracted through the measurement of a set of superallowed nuclear $\beta$ decay;
\item $V_{us}$,  $V_{ub}$ and $V_{cs}$ are measured through the rates of inclusive and exclusive charmless semi-leptonic $K$, $B$ and $D$ decays to $\pi\ell\bar\nu$, respectively;
\item $V_{cb}$ is extracted through the measurement of the $B\to D\ell\bar\nu$ decay;
\item $V_{cs}$ and $V_{tb}$ can be extracted from the measurements of $D\to K\ell\bar\nu$ and top decay to $Wb$, respectively. However, the corresponding constraint is not competitive with the constraint coming from the global fit of all the other observables. 
\item The one loop mass splittings in the neutral $B$ and $B_s$ systems are sensitive to the values of $V_{td}$ and $V_{ts}$, respectively. Additional determinations include loop-mediated rare K and B decays. 
\end{itemize}
$V_{ud}$ is the best determined element of the CKM matrix with an error at the level of $0.02\%$. $V_{us}$, $V_{cs}$, and $V_{cb}$ are also well determined with the corresponding observables with errors ranging in $(0.1-2)\%$. The observables $B\to\pi\ell\bar\nu$ and $D\to\pi\ell\bar\nu$ determining $V_{ub}$ and $V_{cd}$ are, instead, the least accurately measured with an error at around $\sim 10\%$.

 \begin{figure}[t!]
\begin{center}
\includegraphics[width=.5\textwidth]{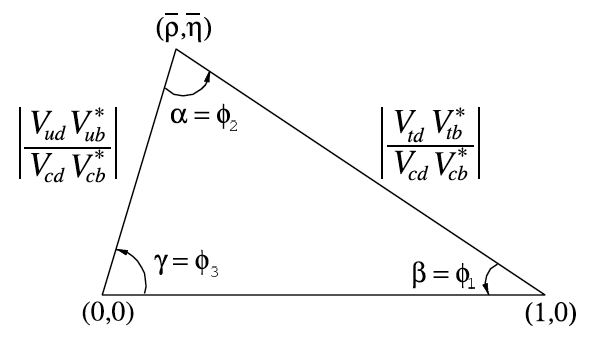}~~
\raisebox{0.2in}{\includegraphics[width=.5\textwidth]{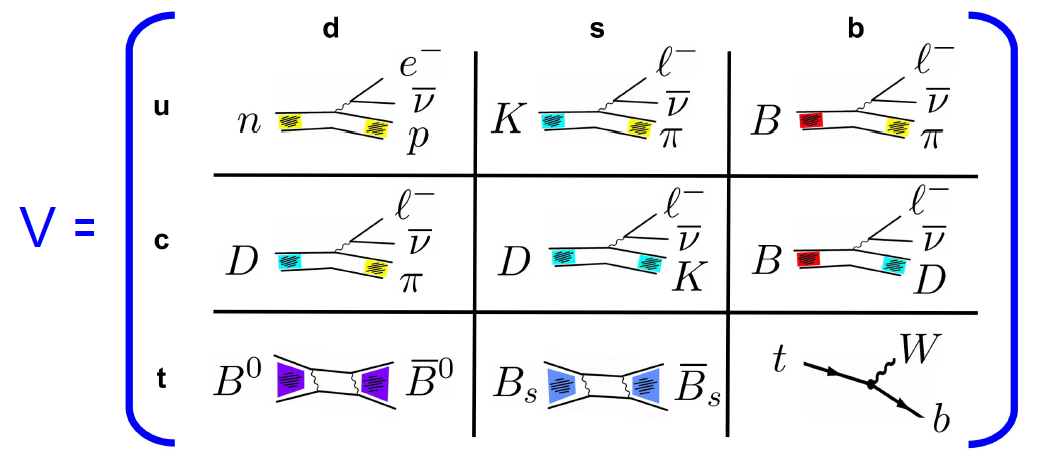}}
\caption{{\bf{Left:}} The unitarity triangle. {\bf{Right:}} List of the most
sensitive observables used to determine the several elements of the CKM matrix.}
\label{fig:CKM}
\end{center}
\end{figure}

The consistency of different constraints on the CKM unitarity triangle is a powerful test of the SM in describing flavor changing phenomena. Fig. \ref{fig:CKMConstraint} shows the huge improvement in the determination of the unitarity triangle in the past 20 years: the left panel shows the present status and the right panel represents the situation back in 1995. In this fit, additional constraints beyond the ones discussed above are imposed. In particular, constraints on the CKM unitarity triangle come from the CP violation in $K\to\pi\pi$, the rates of the various $B\to\pi\pi,\rho\pi,\rho\rho$ decays (that depend on the phase $\alpha$), the rates of various $B\to DK$ decays (that depends on the phase $\gamma$), the CP asymmetry in the decay $B\to\psi K_s$ (that depends on the phase $\beta$).
From the figure, it is evident that there is little room for non-SM contributions in flavor changing transitions. The values of $\bar\rho$ and $\bar\eta$ are determined very accurately\footnote{These numbers are taken from the CKMFitter collaboration \cite{Charles:2004jd}. Similar numbers are obtained by the UTFit collaboration \cite{Bona:2006ah} and by {\tt {latticeaverages.org}} \cite{Laiho:2009eu}.}:
\beq
\bar\rho= 0.150^{+0.012}_{-0.006}, ~ \bar\eta=0.354^{+0.007}_{-0.008},
\eeq
together with the parameters $A$ and $\lambda$:
\beq
A=0.823^{+0.007}_{-0.014}, ~\lambda=0.2254^{+0.0004}_{-0.0003}.
\eeq
One can allow for arbitrary new physics (NP) in one or more flavor changing
processes entering the CKM fit. This is particularly interesting in processes that appear in the SM at the loop-level. Then, one can quantitatively
constrain the size of new physics contributions to processes such as neutral meson mixing. This is what we will discuss in the next section.

 \begin{figure}[t!]
\begin{center}
\includegraphics[width=.49\textwidth]{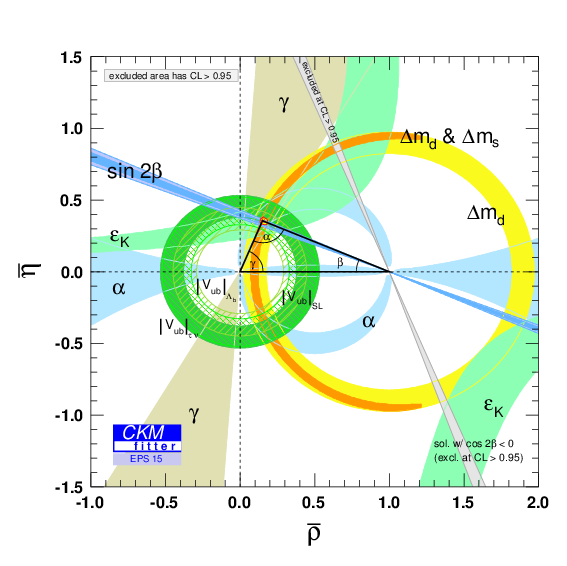}
\includegraphics[width=.48\textwidth]{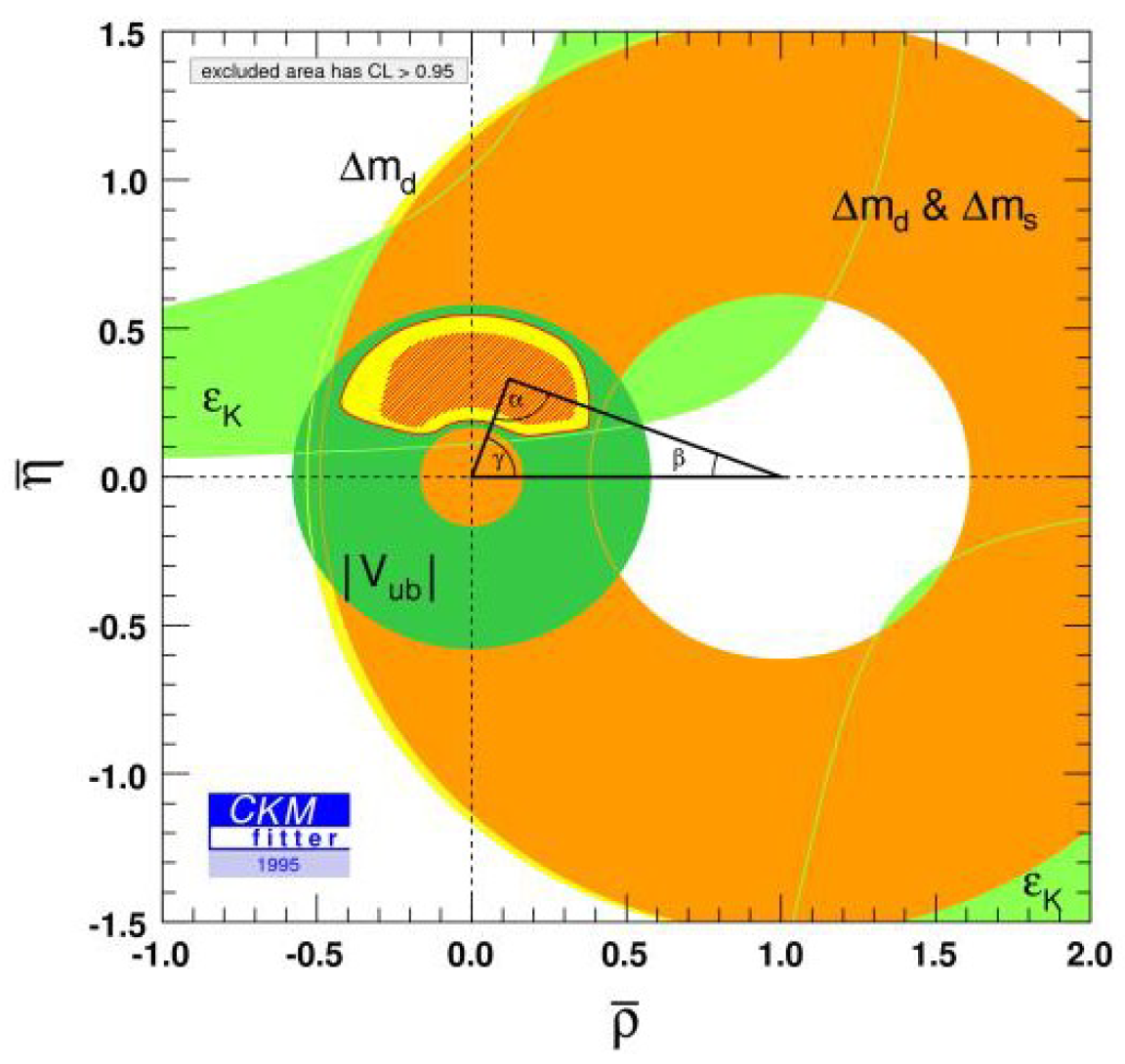}
\caption{Summary of the constraints on the CKM unitarity triangle as obtained by the CKMFitter collaboration \cite{Charles:2004jd} {\bf{Left:}} in 2015, {\bf{Right:}} in 1995.}
\label{fig:CKMConstraint}
\end{center}
\end{figure}

\subsection{Meson mixing and the GIM mechanism}\label{sec:GIMAndKaons}

In the SM, in order for a flavor transition to take place, the
exchange of at least a virtual $W$ is necessary. A {\it{Flavor-Changing-Neutral-Current}} (FCNC) process is a process in which the electric charge does not change between initial and final states. As a consequence, in the SM such processes have a reduced rate relative to a normal weak interaction process. FCNCs are, however, not only suppressed by the loop, but also by the so called Glashow-Iliopoulos-Maiani (GIM) mechanism \cite{Glashow:1970gm}. We will explain this mechanism through the discussion of meson mixing.

Let us take the $K~(=d\bar s)$ and $\bar K~(=\bar d s)$ meson system. These two flavor eigenstates are not mass eigenstates and, therefore, they mix. The leading order contributions to the mixing arise from box diagrams mediated by the exchange of the $W$ boson and the up quarks. The corresponding effective Hamiltonian responsible of this mixing is given by
\beq
\mathcal H_{K}=\frac{G_F^2}{16\pi^2}m_W^2\left[\sum_{i=u,c,t} F(x_i,x_i)\lambda_i^2+\sum_{ij=u,c,t,~i\neq j}F(x_i,x_j)\lambda_i\lambda_j\right](\bar s\gamma_\mu(1-\gamma_5)d)^2,
\eeq
where we have defined $\lambda_i=V_{is}^*V_{id}$. $F(x_i,x_j)$ are loop functions, $x_q\equiv m_q^2/m_W^2$. In the limit of exact flavor symmetry ($m_d=m_s=m_b$) the several diagrams cancel, thanks to the unitarity of the CKM matrix (see Eqs. (\ref{eq:unitarityRelations})). This is the so called {\it{GIM mechanism}}, that can be applied not only to the Kaon mixing system but to all SM flavor transitions. Historically, in 1970, at the time the GIM mechanism was proposed, only three quarks (up, down, and strange) were thought to exist. The GIM mechanism however, required the existence of a fourth quark, the charm, to explain the large suppression of FCNC processes.

The breaking of the flavor symmetry induces a mass difference between the quarks, so the sum of the diagrams responsible for meson mixing will be non-zero. We can use the unitarity relations (\ref{eq:unitarityRelations}) to eliminate the terms in the effective Hamiltonian that depend on $\lambda_u$, obtaining
\beq\label{eq:EffHK}
\mathcal H_{K}=\frac{G_F^2}{16\pi^2}m_W^2\left[S_0(x_t)\lambda_t^2+S_0(x_c)\lambda_c^2+2S_0(x_c,x_t)\lambda_c\lambda_t\right](\bar s\gamma_\mu(1-\gamma_5)d)^2,
\eeq
with $S_0(x_i)$ and $S_0(x_i,x_j)$ given by the combinations 
\bea
S_0(x_i)&\equiv& F(x_i,x_i)+F(x_u,x_u)-2F(x_i,x_u)\\
S_0(x_i,x_j)&\equiv& F(x_i,x_j)+F(x_u,x_u)-F(x_i,x_u)-F(x_j,x_u).
\eea
The explicit expressions can be found in e.g. \cite{Buras:1998raa}. All terms of this effective Hamiltonian are suppressed by, not only the loop factor, but also the small CKM elements, particularly suppressing the top loop contribution, and the small mass ratio $m_c^2/m_W^2$ in the case of the charm loop contribution, as predicted by the GIM mechanism. 

This effective Hamiltonian leads to the oscillation of the two Kaons. The time evolution of the Kaon anti-Kaon system, $\psi=(K,\bar K)$, reads
\beq
i\frac{d\psi(t)}{dt}=\hat H\psi(t), ~~~~\hat H =\hat M-i\frac{\hat\Gamma}{2}=\left(\begin{array}{cc}
M-i\Gamma/2                        & M_{12}-i\Gamma_{12}/2\\
M_{12}^*-i\Gamma_{12}^*/2 & M-i\Gamma/2
\end{array}\right),\eeq
with $M$ and $\Gamma$ the average mass and width of the two Kaons, respectively. The two eigenstates of the system (heavy and light, or, equivalently, long and short)
have a mass and width given by
\bea\nonumber
M_{H,L}&=&M\pm {\rm{Re}}(Q), ~~\Gamma_{H,L}=\Gamma\mp 2 {\rm{Im}}(Q),\\
Q&=&\sqrt{\left(M_{12}-\frac{i}{2}\Gamma_{12}\right)\left(M^*_{12}-\frac{i}{2}\Gamma^*_{12}\right)},
\eea
and are a linear combination of the two $K$ and $\bar K$ states
\beq\label{eq:mixingparam}
|K_{H,L}\rangle =p |K\rangle \mp q |\bar K \rangle, ~~~\frac{q}{p}=-\frac{2M_{12}^*-i\Gamma_{12}^*}{2 {\rm{Re}}(Q)+2i~ {\rm{Im}}(Q)}.
\eeq

The difference in mass of the two Kaon states, $\Delta M_K$, can be computed from the effective Hamiltonian in (\ref{eq:EffHK}) by
\beq\label{eq:dMK}
m_K \Delta M_K=2m_K {\rm{Re}}(M_{12})={\rm{Re}}(\langle \bar K|\mathcal H_{K}|K\rangle),
\eeq
with $m_K$ the average Kaon mass. Lattice QCD is essential to compute the matrix element of the four quark operator calculated between two quark bound states. We have \cite{Buras:1998raa}
\beq
\langle \bar K|(\bar s\gamma_\mu(1-\gamma_5)d)^2|K\rangle=\frac{8}{3}B_K(\mu)F_K^2m_K^2,
\eeq
with $F_K$ the Kaon decay constant and $B_K(\mu)$ the Kaon bag parameter, evaluated at the scale $\mu$.
Putting these pieces together and including QCD corrections, one can find 
\beq
M_{12}=\frac{G_F^2}{12\pi^2}F_K^2\hat B_K m_K m_W^2\left[(\lambda_c^*)^2\eta_1S_0(x_c)+(\lambda_t^*)^2\eta_2S_0(x_t)+2\lambda_c^*\lambda_t^*\eta_3S_0(x_c,x_t)\right],
\eeq
where $\eta_{1,2,3}$ are QCD correction factors given e.g. in \cite{Buras:1998raa} and we have defined the renormalization group invariant parameter
\beq\label{eq:BK}
\hat B_K=B_K(\mu)\left[\alpha_s(\mu)\right]^{-2/9}\left[1+\frac{\alpha_s(\mu)}{4\pi}J_3\right],
\eeq
with $J_3\sim 1.9$ in the NDR-scheme \cite{Buchalla:1995vs}. In Sec. \ref{Sec:EFT}, we will discuss the bounds on New Physics theories arising from the measurement of the several observables of the meson mixing systems.

\subsection{CP violation in meson decays}
All CP-violating observables in $K$ and $\bar K$ decays, as well as in any $M-\bar M$ meson system, to final states $f$ and $\bar f$ can be expressed in terms of phase-convention-independent combinations of $A_f,\bar A_f, A_{\bar f},\bar A_{\bar f}$, together with $q/p$ of Eq. (\ref{eq:mixingparam}), in the case of neutral-mesons, where we define
\beq
A_f=\langle f|\mathcal H|M\rangle,~\bar A_f=\langle f|\mathcal H|\bar M\rangle,~A_{\bar f}=\langle \bar f|\mathcal H|M\rangle,~\bar A_{\bar f}=\langle \bar f|\mathcal H|\bar M\rangle.\eeq

 \begin{figure}[t!]
\begin{center}
\includegraphics[width=.32\textwidth]{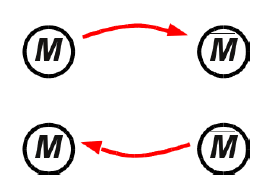}~~
\includegraphics[width=.325\textwidth]{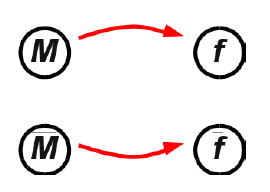}~~\includegraphics[width=.32\textwidth]{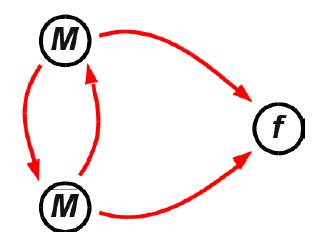}
\caption{{\bf{Left:}} CP violation in mixing; {\bf{Middle:}} CP violation in decay; {\bf{Right:}} CP violation in interference, for meson decays to a final state $f$.}
\label{fig:CPviolation}
\end{center}
\end{figure}

As shown in Fig. \ref{fig:CPviolation}, we distinguish three types of CP-violating effects in meson decays \cite{Nir:2005js}:

\begin{itemize}
\item[(a)] CP violation in mixing, defined by $|q/p|\neq 1$ and arising when the two neutral mass eigenstate admixtures
cannot be chosen to be CP-eigenstates;
\item[(b)] CP violation in the decay of mesons, defined by $|\bar A_{\bar f}/A_f|\neq 1$;
\item[(c)] CP violation in interference between a decay without mixing, $M\to f$, and a decay with mixing $M\to \bar M\to f$. This is defined by ${\rm{Im}}(q\bar A_f/p A_f)\neq 0$.
\end{itemize}

One example of CP violation in mixing (a) is the asymmetry in charged-current semi-leptonic neutral meson decays for which the ``wrong sign" decays (i.e. decays to a lepton of charge opposite to the sign of the charge of the original b quark) are allowed only if there is a mixing between the meson and the anti-meson. For example, for a $B_d$ meson
\beq
a_{\rm{SL}}^d=\frac{\Gamma(\bar B_d(t)\to\ell^+\nu X)-\Gamma(\bar B_d(t)\to\ell^-\bar\nu X)}{\Gamma(\bar B_d(t)\to\ell^+\nu X)+\Gamma(\bar B_d(t)\to\ell^-\bar\nu X)}=\frac{1-|q/p|^4}{1+|q/p|^4}.
\eeq
D0 performed several measurements of these asymmetries in $B$ decays \cite{Abazov:2010hj,Abazov:2009wg,Abazov:2013uma}. Combining all measurements, there is a long-standing anomaly with the SM prediction in the $a_{\rm{SL}}^s-a_{\rm{SL}}^d$ plane with a significance at the level of $\sim2-3\sigma$ \cite{Artuso:2015swg}, mainly arising from the D0 measurement of the like-sign dimuon charge asymmetry \cite{Abazov:2013uma} (see upper panel of Fig. \ref{fig:betabetas}).

In charged meson decays, where mixing effects are absent, the CP violation in decay (b) is the only possible source of
CP asymmetries. For example, in the $B$ meson system:
\beq
a_{f^\pm}=\frac{\Gamma(B^+\to f^+)-\Gamma(B^-\to f^-)}{\Gamma(B^+\to f^+)+\Gamma(B^-\to f^-)}=\frac{1-|\bar A_{f^-}/A_{f^+}|^2}{1+|\bar A_{f^-}/A_{f^+}|^2}.
\eeq
These asymmetries are different from zero only if at least two terms of the amplitude have different weak phases and different strong phases \footnote{Strong phases do not violate CP. Their origin is the contribution from intermediate on-shell states in the decay process, that is an absorptive part of an amplitude.}.
Non-zero CP asymmetries have been observed in a few B meson decay modes by the LHCb collaboration: $B^+\to K^+K^-K^+$, $B^+\to K^+K^-\pi^+$ \cite{Aaij:2014iva}.

 \begin{figure}[t!]
\begin{center}
~~~~~~~~~~~~\includegraphics[width=.55\textwidth]{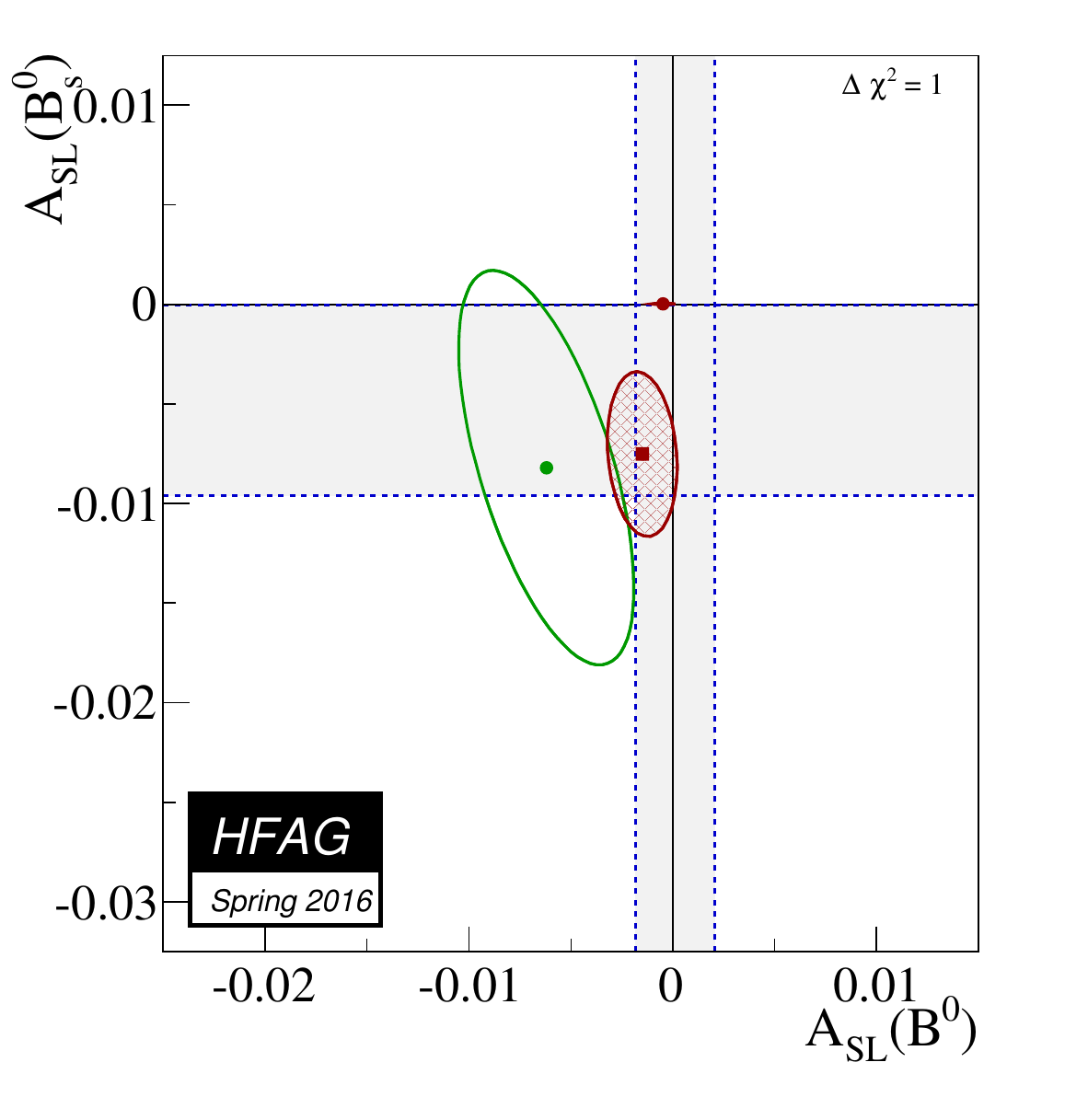}
\includegraphics[width=.6\textwidth]{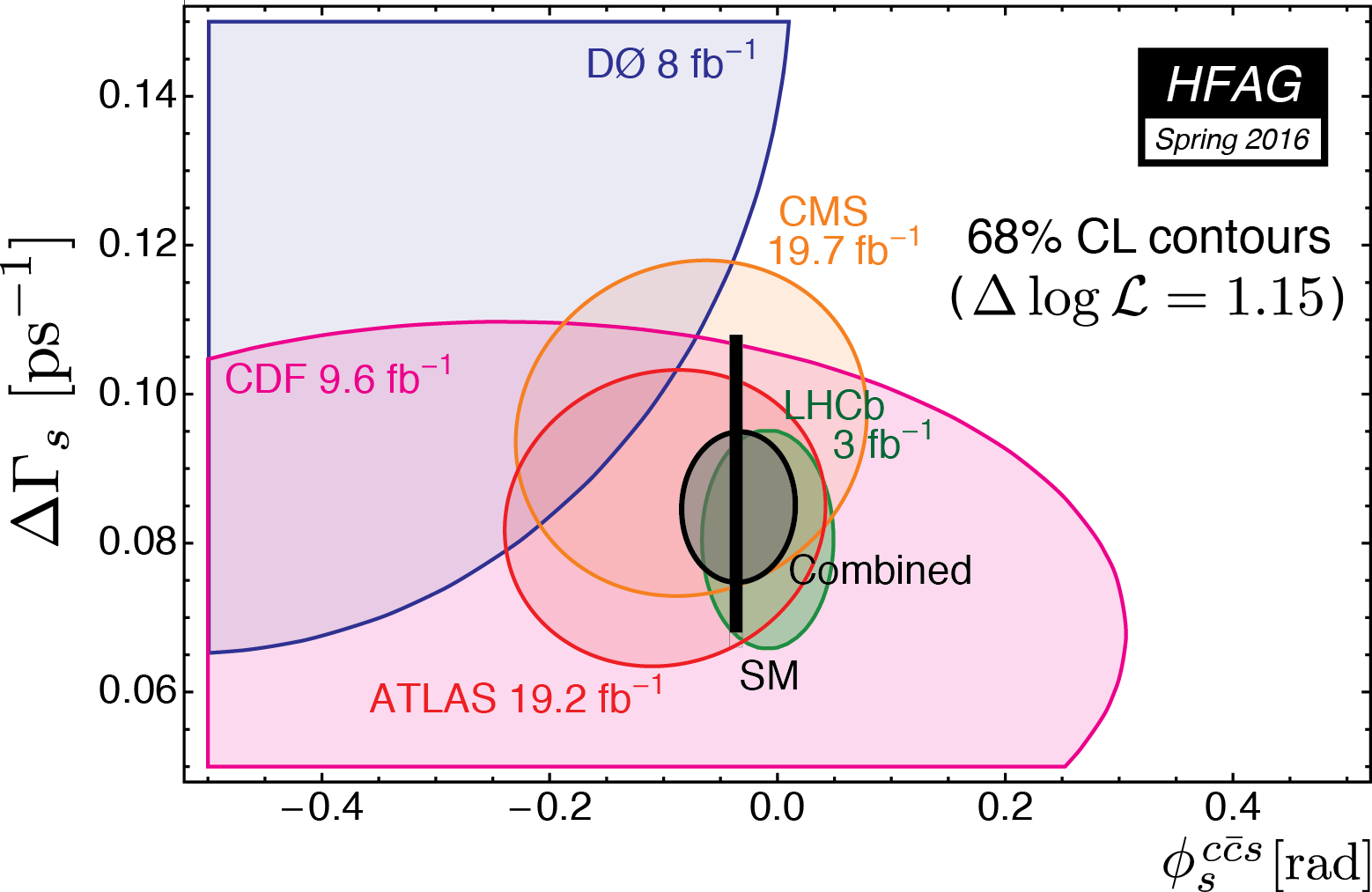}
\caption{{\bf{Upper panel:}} Summary of the measurements of CLEO, BABAR, Belle, D0 and LHCb in the $a_{\rm{SL}}^s-a_{\rm{SL}}^d$ plane. {\bf{Lower panel:}} $(\phi_s^{c\bar c s},\Delta \Gamma_s)$ plane ($\Delta \Gamma_s$ is the difference in width in the $B_s-\bar B_s$ system), the individual $68\%$ confidence-level contours of ATLAS, CMS, CDF, D0 and LHCb, their combined contour (solid line and shaded area), as well as the SM predictions (thin black rectangle) are shown (from \cite{Amhis:2014hma}).}
\label{fig:betabetas}
\end{center}
\end{figure}

CP violation in interference (c) is measured through the decays of neutral mesons and anti-mesons to a final state that is a CP eigenstate ($f_{\rm{CP}}$)
\beq
a_{f_{\rm{CP}}}=\frac{\Gamma(\bar M(t)\to f_{\rm{CP}})-\Gamma(M(t)\to f_{\rm{CP}})}{\Gamma(\bar M(t)\to f_{\rm{CP}})+\Gamma(M(t)\to f_{\rm{CP}})}\simeq \-{\rm{Im}}(\lambda_{\rm{CP}})\sin(\Delta M_M t),\eeq
where we have defined $\lambda_{\rm{CP}}=\frac{q \bar A_f}{p A_f}$ and $\Delta M_M$ is the difference in mass of the meson anti-meson system.
This type of CP violation has been observed in several $B$ meson decays, as for example in $B_d\to J/\psi K_S$ at Babar \cite{Aubert:2009aw}, Belle \cite{Adachi:2012et} and by now by the LHCb, as well \cite{Aaij:2015vza}, leading to the measurement of the $\beta$ angle of the CKM matrix $a_{J/\Psi K_s}\simeq \sin(2\beta)\sin(\Delta M_d t)$. The Feynman diagrams contributing to this asymmetry are given in Fig. \ref{fig:CPInterference}, where we show the tree (left panel) and the penguin (right panel) contributions. 
The current world average on the angle $\beta$ is \cite{Amhis:2014hma}
\beq
\sin(2\beta)=0.69\pm 0.02.
\eeq

The corresponding CP asymmetry in $B_s$ decay is $B_s\to\psi\phi$. The SM prediction is suppressed compared to the $\beta$ angle by $\lambda^2$, leading to $\beta_s^{\rm{SM}}=0.01882^{+0.00036}_{-0.00042}$ \cite{Charles:2004jd}. The latest LHCb result
using 3 fb$^{-1}$ data is in good agreement with this prediction and reads $\beta_s^{\rm{LHCb}}=0.005\pm 0.0195$. A summary of all measurements of the mixing angle in the $B_s-\bar B_s$ system is reported in the lower panel of Fig. \ref{fig:betabetas} and the world average is \cite{Amhis:2014hma}
\beq
\beta_s=-\frac{\phi_s^{c\bar c s}}{2}=-0.00165\pm 0.00165.\eeq

 \begin{figure}[t!]
\begin{center}
\includegraphics[width=.85\textwidth]{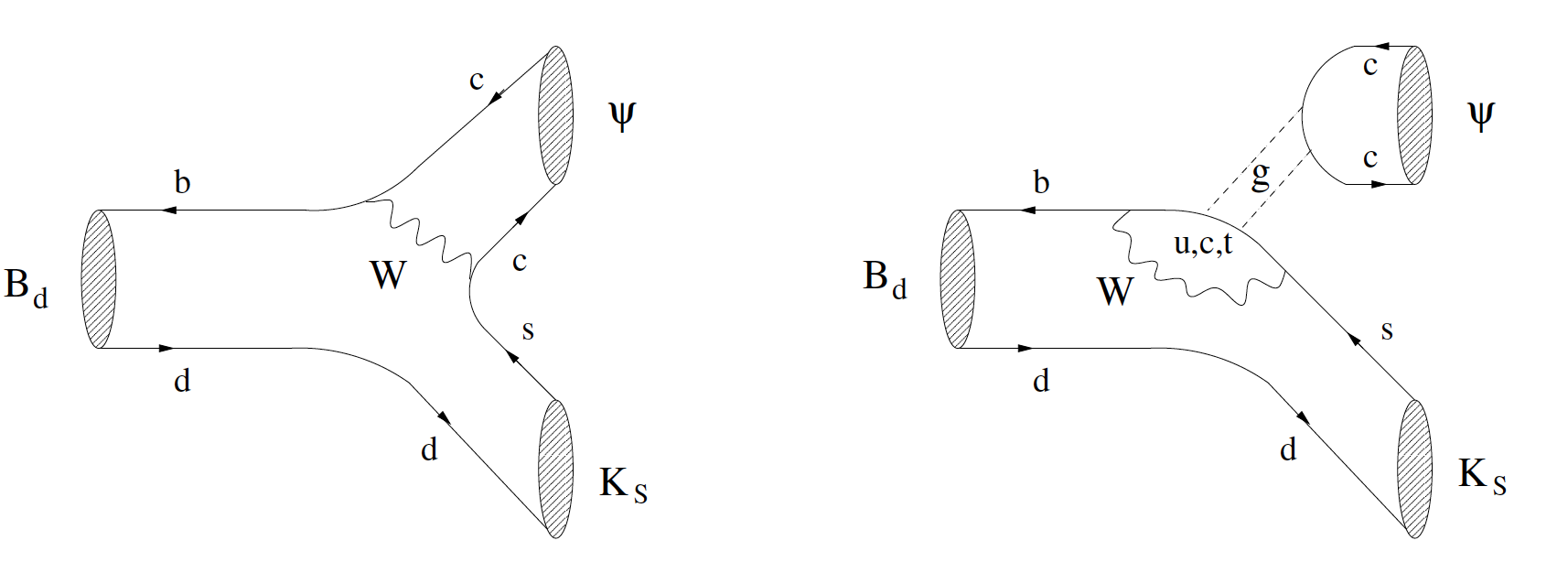}~~
\caption{The {\bf{Left:}} tree diagram and the {\bf{Right:}} penguin diagram contributing to $B_d\to \Psi K_S$ (from \cite{Fleischer:2002ys}).}
\label{fig:CPInterference}
\end{center}
\end{figure}

\section{Effective field theories and  flavor transitions}\label{Sec2}

It is clear that the Standard Model is not a complete theory of Nature. Foremost arguments in favor of the existence of New Physics are 

\begin{itemize}
\item It does not include gravity, and therefore it cannot be valid at energy scales above the Planck scale;
\item It cannot explain the small value of the Higgs boson mass;
\item It cannot account for neutrino masses and for the existence of Dark Matter (DM).
\end{itemize}
In particular, the DM and Higgs mass motivations suggest that the SM should be replaced by a new theory already at scales at around the TeV scale.
Given that the SM is only an effective low energy theory, non-renormalizable terms must be added to the SM Lagrangian. In the next subsection, we will discuss the flavor constraints on the NP scale associated to the higher dimensional operators contributing to flavor transitions.

\subsection{The New Physics flavor puzzle}\label{Sec:EFT}
If we assume that the new degrees of freedom which complete the theory of Nature are heavier than the SM particles, we can integrate them out and describe physics beyond the SM by means of an
effective field theory (EFT) approach. The SM Lagrangian becomes the renormalizable part of this generalized
Lagrangian which includes an infinite sum of operators with dimension $d \geq 5$, constructed in terms
of SM fields and suppressed by inverse powers of the NP scale $\Lambda~(\gg v)$. This approach is a generalization of the Fermi theory of weak interactions, where the dimension six four-fermion operators describing weak decays are the results of having integrated out the $W$ boson. The generic effective Lagrangian reads

\beq\label{eq:effectiveL}
\mathcal L_{\rm{eff}}=\mathcal L_{\rm{SM}}+\sum^{d\geq 5}_n\frac{c_n^d}{\Lambda^{d-4}}\mathcal O_n^{(d)}({\rm{SM}}),
\eeq
where $\mathcal L_{\rm{SM}}$ is the sum of (\ref{eq:LgaugeH}) and (\ref{eq:Yukawas}) and $\mathcal O_n^{(d)}({\rm{SM}})$ are operators of dimension $d \geq 5$ containing SM fields
only and compatible with the SM gauge symmetry. Generically, we would expect the Wilson coefficients $c_n^d=\mathcal O(1)$, however several of these operators contribute to flavor-changing processes and should be very suppressed to be in agreement with low energy flavor experiments. This is often denoted as the {\it{NP flavor puzzle}}.

As an example, we consider the dimension 6 operators contributing to Kaon mixing:
\bea\nonumber
O_1^{\rm{VLL}}&=&(\bar s\gamma_\mu P_L d)^2,\\\nonumber
O_1^{\rm{LR}}&=&(\bar s\gamma_\mu P_L d)(\bar s\gamma^\mu P_R d),\\\label{eq:dim6ope}
O_2^{\rm{LR}}&=&(\bar s P_L d)(\bar s P_R d),\\\nonumber
O_1^{\rm{SLL}}&=&(\bar s P_L d)^2,\\\nonumber
O_2^{\rm{SLL}}&=&(\bar s\gamma_{\mu\nu} P_L d)(\bar s\gamma^{\mu\nu} P_L d),
\eea
plus the corresponding ones with the exchange $P_L\to P_R$ ($P_{L,R}=(1\mp\gamma_5)/2$). The only operator that arises in the SM is $O_1^{\rm{VLL}}$ (see Sec. \ref{sec:GIMAndKaons}). As an example, a NP toy model containing a TeV scale new $Z^\prime$ gauge boson with coupling $g^\prime Z_\mu^\prime(\bar s \gamma^\mu(1-\gamma_5)d)$ would produce a contribution to the operator $O_1^{\rm{VLL}}$ and, therefore, to the difference in mass of Kaon and anti-Kaon system that is equal to\footnote{{\underline{{\bf{Exercise:}}}} compute this contribution.}
\beq
\Delta M_K=\Delta M_K^{\rm{SM}}+\frac{8}{3}m_KF_K^2\hat B_K\frac{(g^\prime)^2}{m_{Z^\prime}^2},\eeq
where $\Delta M_K^{\rm{SM}}$ is the value predicted by the SM, as reported in Eqs. (\ref{eq:dMK})-(\ref{eq:BK}). For TeV-scale $Z^\prime$s coupled to a bottom and a strange quark with a EW strength coupling, the second piece of this equation is $\sim 4$ orders of magnitude larger than the SM contribution, and therefore, such gauge bosons are completely ruled out by Kaon mixing measurements. This shows the tension between a generic NP at around the TeV scale with EW-strength flavor violating couplings and low energy flavor measurements, the so called {\it{NP flavor puzzle}}. 

A summary of the bounds for the four neutral meson systems ($K,B_d,B_s,D$) is shown in Table \ref{tab:DF2}. Particularly, we show in the first two entries the bounds on the NP scale, $\Lambda$, having fixed the absolute value of the corresponding Wilson coefficient, $c_n^6$ of Eq (\ref{eq:effectiveL}), to one (the first column is for $c_n^6=1$, the second one for $c_n^6=i$); the last two columns represent, instead, the bound on real part and on the imaginary part of the the Wilson coefficient, fixing the NP scale to 1 TeV. A few comments are in order. The bounds are weakest (strongest) for $B_s$ ($K$) mesons, as mixing is the least (most) suppressed in the SM in that case. The bounds on the operators with a different chirality (left-right (LR) or right-left (RL)) are stronger, especially in the Kaon case, because of the larger hadronic matrix elements.
Throughout the table, bounds on the NP scale $\Lambda$ exceed the TeV scale by several orders of magnitude. Therefore, we can conclude that, if NP exists at around the TeV scale, it has to possess a highly non-generic flavor structure, to explain $c_n^d\ll 1$.

\begin{table}[t]
\begin{center}
\renewcommand{\arraystretch}{1.2}
\begin{tabular}{||c|c c|c c||} \hline\hline
\rule{0pt}{1.2em}%
Operator &  \multicolumn{2}{c|}{Bounds on $\Lambda$~in~TeV~($c_n^6=1$)} &
\multicolumn{2}{c|}{Bounds on
$c_n^6$~($\Lambda=1$~TeV) }\cr
&   Re& Im & Re & Im   \cr
 \hline\hline $(\bar s_L \gamma^\mu d_L )^2$  &~$9.8 \times 10^{2}$& $1.6 \times 10^{4}$ 
&$9.0 \times 10^{-7}$& $3.4 \times 10^{-9}$  \\ 
($\bar s_R\, d_L)(\bar s_L d_R$)   & $1.8 \times 10^{4}$& $3.2 \times 10^{5}$ 
&$6.9 \times 10^{-9}$& $2.6 \times 10^{-11}$  \\ 
 \hline $(\bar c_L \gamma^\mu u_L )^2$  &$1.2 \times 10^{3}$& $2.9 \times 10^{3}$ 
&$5.6 \times 10^{-7}$& $1.0 \times 10^{-7}$  \\ 
($\bar c_R\, u_L)(\bar c_L u_R$)   & $6.2 \times 10^{3}$& $1.5 \times 10^{4}$ 
&$5.7 \times 10^{-8}$& $1.1 \times 10^{-8}$ \\ 
\hline$(\bar b_L \gamma^\mu d_L )^2$    &  $6.6 \times 10^{2}$ & $ 9.3 \times 10^{2}$ 
&  $2.3 \times 10^{-6}$ &
$1.1 \times 10^{-6}$ \\ 
($\bar b_R\, d_L)(\bar b_L d_R)$  &   $  2.5 \times 10^{3}$ & $ 3.6
\times 10^{3}$ &  $ 3.9 \times 10^{-7}$ &   $ 1.9 \times 10^{-7}$ 
 \\
\hline $(\bar b_L \gamma^\mu s_L )^2$    &  $1.4 \times 10^{2}$ &  $  2.5 \times 10^{2}$   &  
 $5.0 \times 10^{-5}$ &   $1.7 \times 10^{-5}$ 
   \\ 
($\bar b_R \,s_L)(\bar b_L s_R)$  &    $ 4.8  \times 10^{2}$ &  $ 8.3  \times 10^{2}$  & 
   $8.8 \times 10^{-6}$ &   $2.9 \times 10^{-6}$  
   \\ \hline\hline
\end{tabular}
\caption{\label{tab:DF2} Bounds on representative dimension-six operators that mediate meson mixing, assuming an effective coupling $c_n^6/\Lambda^2$.
The bounds quoted for $\Lambda$ are obtained setting 
$|c_n^6|=1$; those for  $c_{\rm NP}$ are obtained setting
$\Lambda=1$ TeV. We define $q_{L,R}\equiv P_{L,R} ~q$. From \cite{Isidori:2010kg} and \cite{Isidori:2013ez}.}
\end{center}
\end{table}

\subsection{The Minimal Flavor Violation ansatz}\label{Sec:MFV}

TeV scale New Physics could be invariant under some flavor symmetry, and, therefore, more easily in agreement with low energy flavor measurements. One example, of a class of such models are theories with Minimal Flavor Violation (MFV) \cite{Chivukula:1987py,Hall:1990ac,Buras:2000dm,D'Ambrosio:2002ex}. Under this assumption, flavor violating
interactions are linked to the known structure of the SM Yukawa couplings also beyond the SM. More specifically, the MFV ansatz can be implemented within the generic effective Lagrangian
in Eq. (\ref{eq:effectiveL}), as well as to UV complete models, and it consists of two ingredients \cite{D'Ambrosio:2002ex}: (i) a flavor symmetry and (ii) a set of symmetry-breaking terms. The symmetry is the SM global symmetry in absence of Yukawa couplings, as shown in Eq. (\ref{eq:flavorSymm}). Since this global symmetry, and
particularly the $SU(3)$ subgroups controlling quark flavor-changing transitions, is broken within the SM, it cannot be promoted to an exact symmetry of the NP model. Particularly, in the SM we can formally recover the flavor invariance under $\mathcal G_{\rm{flavor}}$ by promoting the Yukawa couplings $Y_d,Y_u,Y_e$ of (\ref{eq:Yukawas}) to dimensionless auxiliary fields (spurions) transforming under $SU(3)^3_q=SU(3)_Q\times SU(3)_U\times SU(3)_D$ and under $SU(3)^2_\ell=SU(3)_L\times SU(3)_e$ as 
\beq\label{eq:YukawaTransformations}
Y_Q\sim(3,1,\bar 3)_{SU(3)^3_q},~Y_u\sim(3,\bar 3,1)_{SU(3)^3_q},~Y_e\sim (3,\bar 3)_{SU(3)^2_\ell}.\eeq
{\underline{{\bf{Exercise:}}}} Check that, with these transformations, the Yukawa Lagrangian of (\ref{eq:flavorSymm}) is invariant under $SU(3)^3_q\times SU(3)^2_\ell$.

Employing an effective field theory language, a theory satisfies the MFV ansatz, if all higher-dimensional operators, constructed from SM and $Y_{u,d,e}$ fields, are invariant under the flavor group, $\mathcal G_{\rm{flavor}}$. The invariance under CP of the NP operators may or may not be
imposed in addition to this criterion. In the down quark sector, the several operators will be combinations of the invariants
\beq
\bar Q_L Y_u Y_u^\dagger Q_L,~\bar D_RY_d^\dagger Y_u Y_u^\dagger Q_L,~\bar D_R Y_d^\dagger Y_u Y_u^\dagger Y_d D_R.\eeq
As an example, let us take the operators in (\ref{eq:dim6ope}) and impose the MFV hypothesis. 
The corresponding Wilson coefficients cannot be generic order one numbers, since the operators are not invariant under the flavor symmetry $\mathcal G_{\rm{flavor}}$. The leading term for the first operator reads
\beq
(c_1^{\rm{VLL}})_{\rm{MFV}}\mathcal O_1^{\rm{VLL}}=Z y_t^4 (V_{ts}^*V_{td})^2(\bar s\gamma_\mu P_L d)^2,
\eeq
where $y_t$ is the SM top Yukawa ($=m_t/v$) and $Z$ is a (flavor independent) coefficient, generically of $\mathcal O(1)$. Thanks to the suppression by the small CKM elements $V_{ts}$ and $V_{td}$, the bound on the NP scale $\Lambda$ of this operator is relatively weak $\Lambda \gtrsim 5$ TeV, to be compared to the bound of $1.6\times 10^4$ TeV, as shown in Tab. \ref{tab:DF2}.
The other operators have, instead, a much smaller Wilson coefficient as they are suppressed by either the strange Yukawa square ($\mathcal O_1^{\rm{SLL}},~\mathcal O_2^{\rm{SLL}}$) or the product of down and strange Yukawas ($\mathcal O_1^{\rm{LR}},~\mathcal O_2^{\rm{LR}}$), resulting also in weak bounds on the NP scale $\Lambda$. {\underline{{\bf{Exercise:}}}} write the leading term of the Wilson coefficient of each operator in (\ref{eq:dim6ope}), according to the MFV ansatz and demonstrate that they are much smaller than $(c_1^{\rm{VLL}})_{\rm{MFV}}$.

This structure can be generalized to any higher dimensional operator mediating a flavor transition. Thus, generically in MFV models, 
flavor changing operators automatically have their SM-like suppressions, proportional to the same CKM elements and quark masses as in the SM and this can naturally address the NP flavor puzzle, as the NP scale of MFV models
can be $\mathcal O(1~{\rm{TeV}})$ without violating 
flavor physics bounds. 

To conclude, the MFV ansatz is remarkably successful in satisfying the constraints from low energy flavor observables. However, it does not address the question {\it{Why do quark and lepton masses, as well as quark mixing, have such a hierarchical pattern}} (SM flavor puzzle), since it simply states that the NP flavor violation has to have the same structure of the SM flavor violation.

\subsection{\boldmath Effective field theories for rare $B$ decays}
Rare $B_d$ and $B_s$ decays based on the $b \to s$ flavor changing neutral-current transition are very sensitive to BSM, as they are very suppressed in the SM \cite{Altmannshofer:2013oia}. In the last few years, measurements at the LHC, complementing earlier B-factory results, have hugely increased the available experimental information on these decays. In these lectures, we will focus on the golden channels: the $B_s$ and $B_d$ decays to two muons as they are among the rarest $B$ decays. 
(see \cite{Blake:2016olu} for a recent review, that discusses additional $B$ rare decays, as for example $B_s\to K\mu^+\mu^-$ and $B_s\to K^*\mu^+\mu^-$). 

In the SM, these decays are dominated by the $Z$ penguin and box diagrams involving top quark exchanges. The resulting effective Hamiltonian depends, therefore, on the loop function $Y(x_t)$ (see e.g. \cite{Altmannshofer:2009ne} for its definition), with $x_t\equiv m_t^2/m_W^2$ and reads
\beq\label{eq:HeffBmumuSM}
\mathcal H_{\rm{eff}}=-\frac{G_F}{\sqrt 2}\frac{\alpha}{\pi\sin^2\theta}V_{tb}^*V_{ts}Y(x_t)(\bar b\gamma_\mu P_Ls)(\bar \mu\gamma_\mu\gamma_5\mu)+{\rm{h.c.}},
\eeq
with $s$ replaced by $d$ in the case of $B_d\to\mu^+\mu^-$. Evaluating the two matrix elements of the quark current and of the muon current leads to the branching ratio
\beq\label{eq:BRmumuSM}
{\rm{BR}}(B_s\to\mu^+\mu^-)=\frac{G_F^2}{\pi}\left(\frac{\alpha}{4\pi\sin^2\theta}\right)^2|V_{tb}^*V_{ts}|^2Y^2(x_t)m_\mu^2m_{B_s}\sqrt{1-\frac{4m_\mu^2}{m_{B_s}^2}}F_{B_s}\tau_{B_s},\eeq
and analogously for the $B_d$ decay. In this equation, $m_{B_s}$ is the mass of the $B_s$ meson, $\tau_{B_s}$ its life time (1.6 ps), and $F_{B_s}$ the corresponding decay constant. The main theoretical uncertainties in this branching ratio result from the uncertainties in the decay constant ($\sim4\%$ for $B_d$ and $\sim 3\%$ for $B_s$, using the latest lattice computations \cite{Aoki:2016frl}), and in the CKM elements $V_{td}$ and $V_{ts}$ (both at the level of several $\%$ \cite{Charles:2004jd}). Inserting numbers and including the $\mathcal O(\alpha)$ and $\mathcal O(\alpha_s^2)$ corrections, the latest SM predictions read \cite{Bobeth:2013uxa}
\beq\label{eq:SMBmumu}
{\rm{BR}}(B_s\to\mu^+\mu^-)_{\rm{SM}}=(3.65\pm 0.23)\times 10^{-9},~~~{\rm{BR}}(B_d\to\mu^+\mu^-)_{\rm{SM}}=(1.06\pm 0.09)\times 10^{-10}.
\eeq
As shown by Eq. (\ref{eq:BRmumuSM}), the tiny branching ratios of these decays in the SM are due to several factors: (i) loop suppression, (ii) CKM suppression, and (iii) helicity suppression (by the small muon mass, $m_\mu$). As we will discuss later in this section, extensions of the SM do not necessarily contain any of these suppression mechanisms, and, more in particular, the helicity suppression (iii).

Experimentally, searches for $B_{s,d}\to\mu^+\mu^-$ have been performed by 11 experiments, spanning more than three decades (see upper panel of Fig. \ref{fig:Bmumu} for a summary of all bounds and measurements). In the figure, markers without error bars denote upper limits on the branching fractions at $90\%$ confidence level, while measurements are denoted with error bars delimiting $68\%$ confidence intervals. The first hint for a non-zero $B_s$ decay was reported in 2011 by the CDF collaboration \cite{Aaltonen:2013as}: ${\rm{BR}}(B_s\to\mu^+\mu^-)_{\rm{CDF}}=1.3^{+0.9}_{-0.7}\times 10^{-8}$.  This was followed by several measurements by ATLAS, CMS and LHCb and by the first evidence for a non-zero $B_d$ decay, as observed by the combination of CMS and LHCb Run I analyses \cite{CMS:2014xfa}: ${\rm{BR}}(B_d\to\mu^+\mu^-)_{\rm{CMS+LHCb}}=3.9^{+1.6}_{-1.4}\times 10^{-10}$. In the lower panel of Fig. \ref{fig:Bmumu}, we show the latest status of the measurement of the $B_s$ and $B_d$ decay mode. Particularly, by now, we have a $6.2\sigma$ evidence for $B_s\to\mu^+\mu^-$ with
\beq\label{eq:measurementsBmumu}
{\rm{BR}}(B_s\to\mu^+\mu^-)_{\rm{CMS+LHCb}}=2.8^{+0.7}_{-0.6}\times 10^{-9},~~{\rm{BR}}(B_s\to\mu^+\mu^-)_{\rm{ATLAS}}=0.9^{+1.1}_{-0.8}\times 10^{-9},\eeq
showing a good agreement with the SM prediction  (see \cite{Aaboud:2016ire} for the ATLAS analysis).

 \begin{figure}[t!]
\begin{center}
\includegraphics[width=.69\textwidth]{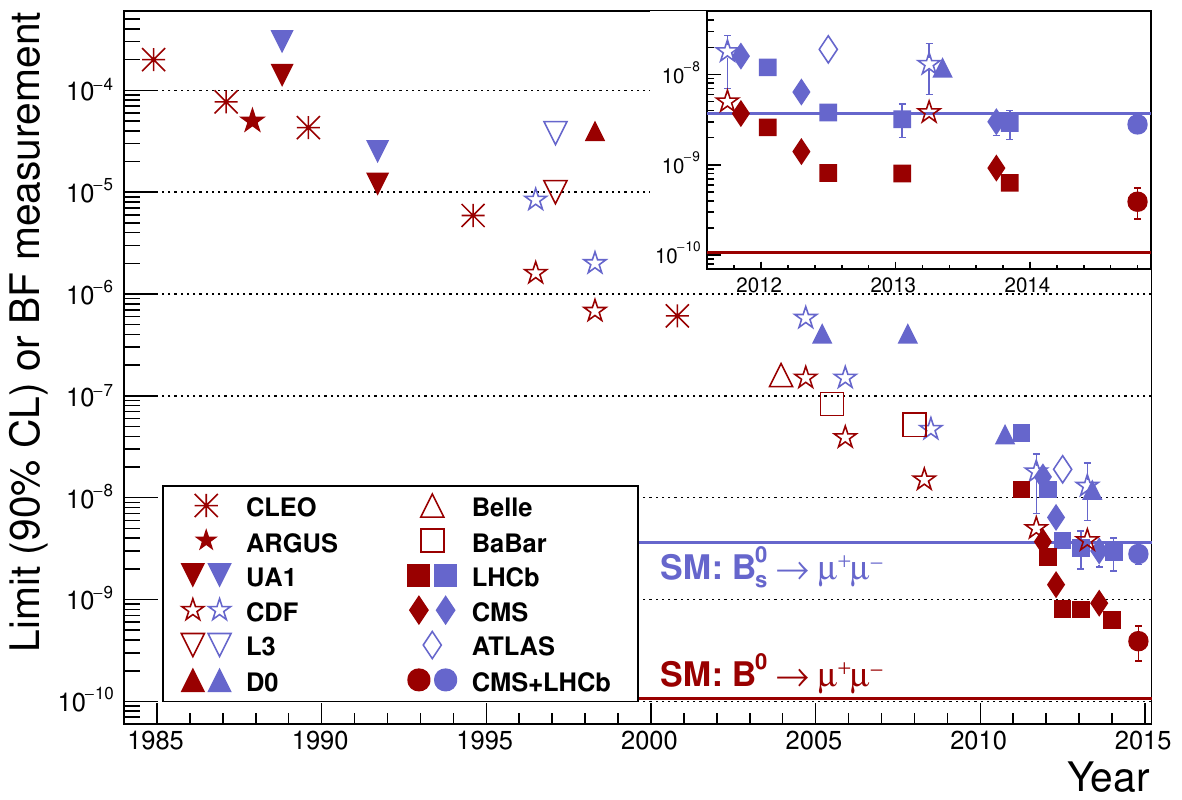}
\includegraphics[width=.69\textwidth]{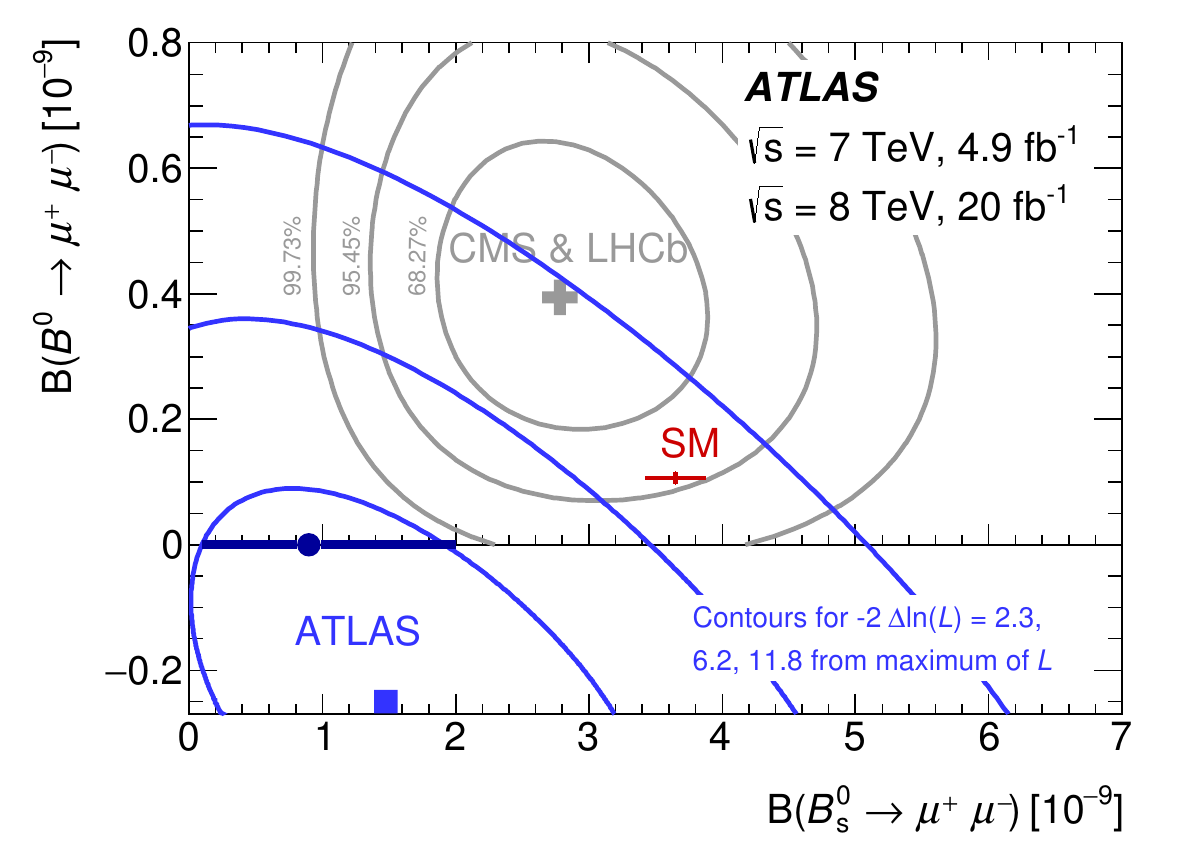}
\caption{{\bf{Upper panel:}} Searches for $B_{s,d}\to\mu^+\mu^-$ from 1985 to 2015. Markers without error bars denote upper limits on the branching fractions at $90\%$ confidence level, while measurements are denoted with errors bars delimiting $68\%$ confidence intervals. The horizontal lines represent the SM predictions for the $B_s\to\mu^+\mu^-$ and $B_d\to\mu^+\mu^-$ branching fractions (from \cite{CMS:2014xfa}); {\bf{Lower panel:}} Present status of the measurements of $B_{s,d}\to\mu^+\mu^-$ at the $1,2,3\sigma$ contours. Shown are the corresponding contours for the combined result of the CMS and LHCb experiments, the ATLAS measurement, and the SM prediction (from \cite{Aaboud:2016ire}).}
\label{fig:Bmumu}
\end{center}
\end{figure}

In BSM theories, several additional operators can contribute to the $B_{s,d}$ decays: $\mathcal O_{10}^\prime$, obtained from the SM operator in (\ref{eq:HeffBmumuSM}) with $P_L\to P_R$ and
\bea\nonumber
\mathcal O_S&=&(\bar bP_L s)(\bar\mu\mu),\\
\mathcal O_P&=&(\bar bP_L s)(\bar\mu\gamma_5\mu),
\eea
and the corresponding prime operators obtained by $P_L\to P_R$. Using these additional operators, one can compute the branching ratio \cite{Altmannshofer:2012az}
\bea
\frac{{\rm BR}(B_s \rightarrow \mu^+ \mu^-)}{{\rm BR}(B_s \rightarrow \mu^+ \mu^-)_{\rm{SM}}} &\simeq& \big(|S_s|^2 + |P_s|^2\big) \nonumber \\
&&\times\bigg(1 + y_s \frac{{\rm Re}(P_s^2) - {\rm Re}(S_s^2)}{|S_s|^2 + |P_s|^2}\bigg)\bigg(\frac{1}{1+y_s}\bigg), \label{eq:bmumuGeneral}
\eea
where $y_s = (8.8 \pm 1.4)\%$  ($y_d \sim 0$ for the $B_d$ system) have to be taken into account when comparing experimental and theoretical results, and
\bea
S_s &\equiv& \frac{m_{B_s}}{2m_\mu}\frac{(C_s^S - C_s^{\prime S})}{C^{SM}_{10\,{s,d}}}\sqrt{1-\frac{4m^2_\mu}{m^2_{B_s}}}, \\
P_s &\equiv& \frac{m_{B_s}}{2m_\mu} \frac{(C_s^P - C_s^{\prime P})}{C^{SM}_{10\,{s,d}}} + \frac{(C_s^{10}-C^\prime_{10\,s})}{C^{SM}_{10\,s}},
\eea
with the several Wilson coefficients defined using the normalization
\beq
\mathcal H_{\rm{eff}}=-4\frac{G_F}{\sqrt 2} V_{tb}V_{ts}^*\frac{e^2}{16\pi^2}\sum _i (C_i \mathcal O_i+C_i^\prime \mathcal O_i^\prime)+{\rm{h.c.}}.
\eeq
Similar expressions hold for the $B_d$ system.
It is evident that the helicity suppression of the branching ratio can be eliminated thanks to the scalar and pseudoscalar operators and, therefore, large enhancements can be obtained. Comparing with the latest measurement of $B_s\to\mu^+\mu^-$, one can find the bounds on the Wilson coefficients of the scalar and pseudoscalar operators, as shown in Fig. \ref{fig:BsmumuConstraint}. The Wilson coefficients of the scalar operators are strongly constrained by the measurement of the $B_s$ rare decay with a bound at the level of Re$(C_s^S-C_s^{\prime S})<0.071$ and Im$(C_s^S-C_s^{\prime S})<0.065$. Scalar NP contributions always increase the branching ratio and, for this reason, the $1\sigma$ region does not appear in the left panel of Fig. \ref{fig:BsmumuConstraint} (the present measurement is smaller than the SM prediction at the $1\sigma$ level, see Eqs. (\ref{eq:SMBmumu}) and (\ref{eq:measurementsBmumu})).
The pseudoscalar Wilson coefficients are instead more weakly constrained (see right panel of Fig. \ref{fig:BsmumuConstraint}), and are consistent with 0 at the $1\sigma$ level. Scalar and pseudoscalar Wilson coefficients for the $B_d$ meson decay are only weakly constrained, at the level of $\mathcal O(0.5)$. 

 \begin{figure}[t!]
\begin{center}
\raisebox{0.05in}{\includegraphics[width=.485\textwidth]{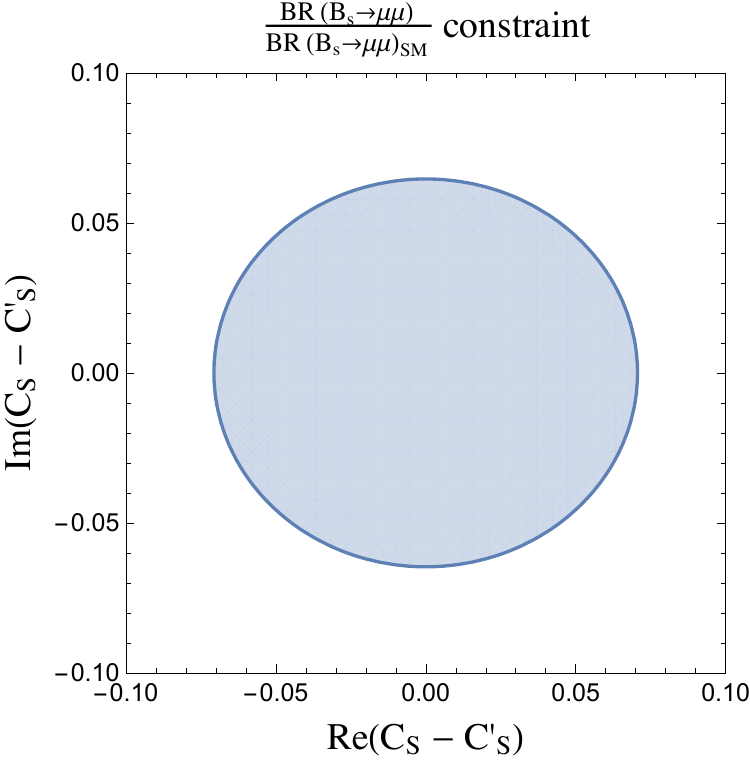}}~~
\includegraphics[width=.48\textwidth]{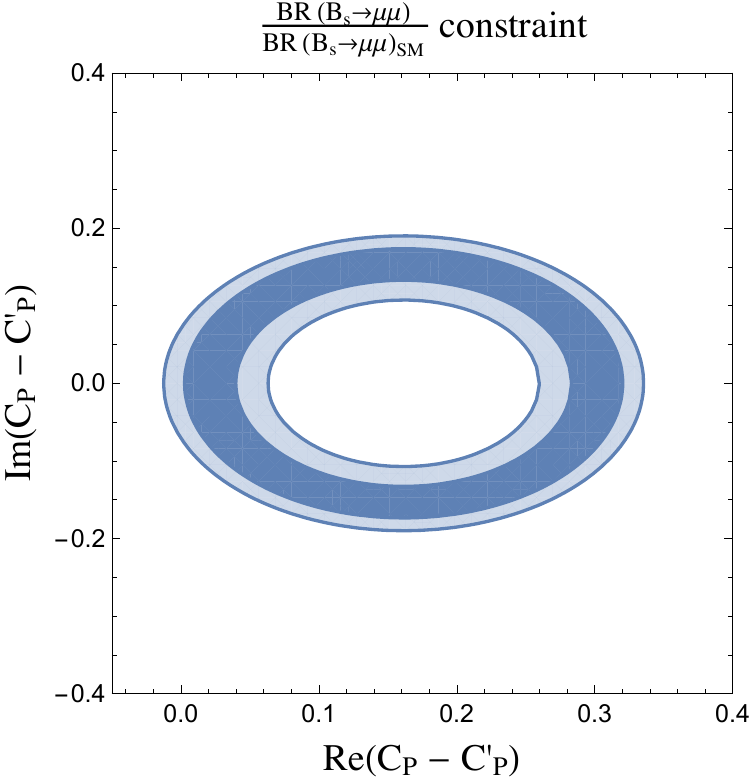}
\caption{One (dark blue) and two (light blue) $\sigma$ bounds on the Wilson coefficients of the {\bf{Left:}} scalar operator and {\bf{Right:}} pseudoscalar operator, as obtained using the latest measurement of $B_s\to\mu^+\mu^-$, assuming no new physics in $C_{10}-C_{10}^\prime$ and switching on one set of operators at a time. Note the change in the axis range in the two panels.}
\label{fig:BsmumuConstraint}
\end{center}
\end{figure}

As it is well known, the measurement of the ratio beween BR$(B_s \to \mu^+\mu^-)$ and BR$(B_d \to \mu^+\mu^-)$ gives a very clean probe of new sources of flavor violation beyond the CKM matrix. Indeed, in all MFV models (see Sec. \ref{sec:MFV2HDM} of these lectures and e.g. \cite{Buras:2010mh} for some examples of MFV models), the ratio is determined by \cite{Buras:2003td}
\beq\label{eq:ratioBmumuMFV}
\frac{{\rm{BR}}(B_d \to \mu^+\mu^-)}{{\rm{BR}}(B_s \to \mu^+\mu^-)}=\frac{\tau_{B_d}}{\tau_{B_s}}\frac{m_{B_d}}{m_{B_s}}\frac{F_{B_d}^2}{F_{B_s}^2}\frac{|V_{td}|^2}{|V_{ts}|^2}\sim 0.03,
\eeq
and has a relatively small theoretical uncertainty at the level of $\sim 5\%$. Presently, the measurement of this ratio by CMS and LHCb is given by $0.14 \pm 0.05$. In the coming years, the LHCb, ATLAS and CMS collaborations will be able to produce a more accurate test of this relation and, therefore, of the MFV ansatz. More specifically, the LHCb upgrade (50 fb$^{-1}$ data) will measure the SM prediction of this ratio with an uncertainty of $\sim 35\%$ \cite{Bediaga:2012py}.

\section{Flavor at high energy: NP models and predictions} \label{Sec3}

In this section we discuss the synergy between {\it{direct searches}} for NP particles at the LHC and {\it{indirect searches}} for NP through the measurement of flavor transitions at B-factories and at the LHCb. We will focus on specific NP frameworks: Two Higgs doublet models (2HDMs) in Sec. \ref{sec:MFV2HDM} and Supersymmetric (SUSY) models in Sec. \ref{sec:SUSY}, with new particles with masses at around the EW scale, that generically can not be integrated out to match the effective theories presented in the previous section. Historically, a few particles have been discovered first indirectly.  In 1970, the measurement of the tiny
branching ratio for the decay $K_L\to\mu^+\mu^-$ lead to the prediction of the existence of the charm quark by Glashow, Iliopoulos and Maiani, before the direct discovery of the $J/\Psi$ charm meson in 1974 by SLAC and BNL. Another remarkable example was the observation of CP violation in Kaon anti-Kaon oscillations that lead to the prediction of the existence of a third generation quarks by Kobayashi and Maskawa in 1973. The direct discovery of the bottom quark came four years later at Tevatron.

\subsection{A Two Higgs doublet model with MFV}\label{sec:MFV2HDM}
Two Higgs doublet models arise in several extensions of the SM, as for example Supersymmetric models. 
In the presence of more than one Higgs field the appearance of tree-level FCNC is not automatically forbidden by the GIM mechanism: additional conditions \cite{Glashow:1976nt,Paschos:1976ay} have to be imposed on the model in order to guarantee a sufficient suppression of FCNC processes. The most general 2HDM has, in fact, several new sources of flavor and of CP violation. Particularly, the Higgs potential is given by\footnote{See \cite{Branco:2011iw} for a review about 2HDMs.}
\begin{eqnarray}\nonumber 
	V(H_1,H_2)&=&\mu_1^2|H_1|^2+\mu_2^2|H_2|^2+(b H_1 H_2+{\rm h.c})+\frac{\lambda_1}{2}|H_1|^4+\frac{\lambda_2}{2}|H_2|^4+\lambda_3|H_1|^2|H_2|^2\\
	&+&\lambda_4|H_1H_2|^2+\left[\frac{\lambda_5}{2}(H_1H_2)^2+\lambda_6|H_1|^2H_1H_2+\lambda_7|H_2|^2H_1H_2+{\rm h.c}\right]\,,
	\end{eqnarray}
where $H_1 H_2= H_1^T (i \sigma_2) H_2$. New sources of CP violation can arise from the terms $(H_1H_2)^2,~|H_1|^2H_1H_2$ and $|H_2|^2H_1H_2$, since, in all generality, $\lambda_{5,6,7}$ are complex coefficients. The most general Yukawa interaction Lagrangian can be written as 
\begin{eqnarray}
- \mathcal L_Y^{\rm gen} = \bar Q_L X_{d1} D_R H_1 + \bar Q_L X_{u1} U_R H_1^c 
+ \bar Q_L X_{d2} D_R H_2^c + \bar Q_L X_{u2} U_R H_2 +{\rm h.c.}~,
\label{eq:generalcouplings}
\end{eqnarray}
to which we can add the corresponding terms for the charged leptons (with $X_{\ell 1},X_{\ell 2}$ Yukawas). After EWSB, quarks acquire mass from both $H_1$ ($\langle H_1\rangle=v\cos\beta$) and $H_2$ ($\langle H_2\rangle=v\sin\beta$).
For generic $X_i$ we cannot diagonalize simultaneously the two mass matrices:
\beq
M_i=\frac{v}{\sqrt 2}(\cos\beta X_{i1}+\sin\beta X_{i2}),~~(i=u,d)
\eeq
and the couplings to the additional physical neutral Higgs fields, $H,A$, which are given in the decoupling (or alignment \cite{Gunion:2002zf}) limit, $\cos(\alpha-\beta)=0$, by
\beq
Z_i=\cos\beta X_{i2}-\sin\beta X_{i1},~~(i=u,d)
\eeq
where we have defined the angle $\beta$ as $\tan\beta=v_2/v_1$\footnote{Strictly speaking $\tan\beta$ is not a physical parameter in a generic 2HDM \cite{Davidson:2005cw}, since the two Higgs doublets, $H_1,H_2$, can be
transformed into each other. In the following, we will describe the MFV 2HDM, in which $\tan\beta$ is a well defined quantity.}.
Consequently we are left with dangerous FCNC couplings at the tree-level and with possible additional new sources of CP violation if (some of) the Yukawas are complex. FCNCs at the tree-level can be eliminated by imposing a discrete $Z_2$ symmetry, leading to a Type I, II, X or Y 2HDM \cite{Ferreira:2010xe} or by assuming the proportionality relations $X_{i1}\propto X_{i2}$, as in the {\it{aligned 2HDM}} \cite{Pich:2009sp}.
This alignment condition is, however, not preserved by renormalization group equations, and, therefore, imposing the alignment condition at some high energy scale, as the GUT scale, will not result in an alignment model at the EW scale \cite{withHowie}. 

The MFV ansatz presented in Sec. \ref{Sec:MFV} can be imposed to the 2HDM and this leads to interesting phenomenology both at low \cite{Buras:2010mh} and high energy \cite{Altmannshofer:2012ar}. The four Yukawa couplings $X_{u1},X_{u2},X_{d1},X_{d2}$ will be a combination of the two $Y_u,Y_d$ SM spurions. More specifically, without loss of generality we can define $Y_u,Y_d$ to be the flavor structures appearing in $X_{u2}$ and $X_{d1}$, respectively. Then we can express the two remaining Yukawa interactions as
\bea
X_{d1} &=& Y_d~,  \nonumber \\
X_{d2} &=& \epsilon_{0} Y_d + \epsilon_{1} Y_d^\dagger Y_d Y_d                    
+  \epsilon_{2} Y_u^\dagger Y_u Y_d + \ldots~,   \nonumber \\
X_{u1} &=& \epsilon^\prime_{0} Y_u + \epsilon^\prime_{1} Y_u^\dagger Y_u Y_u 
+  \epsilon^\prime_{2} Y_d^\dagger Y_d Y_u + \ldots~,  \nonumber \\\label{eq:2HDMMFVYukawas}
X_{u2} &=& Y_u~,
\label{eq:XMFVgen} 
\eea
with $\epsilon_i^{(\prime)}$ generic order one (flavor independent) complex coefficients, and where we have suppressed the higher order terms in $Y_d^\dagger Y_d$ and $Y_u^\dagger Y_u$\footnote{See \cite{Kagan:2009bn} for the discussion of the general MFV (GMFV), where both the top and bottom Yukawas are assumed
to be of order one and their effects are re-summed to all orders.}. If the expansions are truncated to the first order, one can recover the alignment condition, $X_{i1}\propto X_{i2}$. However, differently from the alignment model, quantum corrections cannot modify this functional form of the MFV expansion in (\ref{eq:XMFVgen}), but they can only change the values of the $\epsilon_i^{(\prime)}$ at different energy scales. Additionally, for particular choices of the parameters $\epsilon_i^{(\prime)}$ in (\ref{eq:XMFVgen}), one can recover the Type I, II, X and Y 2HDM. {\underline{\bf{Exercise}}: convince your-self that, with the assumption in (\ref{eq:XMFVgen}) and the transformation properties of the Yukawas in Eq. (\ref{eq:YukawaTransformations}), the several Yukawa terms are invariant under the $SU(3)_q^3$ flavor symmetry.

The MFV 2HDM predicts Higgs-mediated FCNCs at the tree-level, arising from the terms $Y_u^\dagger Y_u Y_d$ and $Y_d^\dagger Y_d Y_u$ in (\ref{eq:2HDMMFVYukawas}). However, the flavor changing Higgs couplings are highly non-generic and, as we now discuss, generically leads to FCNCs in agreement with low energy data. Thanks to MFV, the contribution to meson mixing has the same dependence on the quark masses and CKM elements, as in the SM and, e.g. in the case of the difference in mass, reads
\label{eq:M12KHiggsMFV}
\bea
\Delta M_K^{\rm{NP}}&\sim& 2~{\rm{Re}}(M_{12}^K)=\frac{16}{3}M_K F_K^2 P_2^{LR}(K)
\frac{|a_0|^2}{M_H^2} \frac{m_s m_d m_t^4}{v^6} {\rm{Re}}[(V_{ts} V^*_{td})^2]\tan^2\beta~, \\\nonumber
\label{eq:M12dHiggsMFV}
\Delta M_{B_s}^{\rm{NP}}&\sim& 2~|M_{12}^s| = \frac{16}{3}M_{B_s} F_{B_s}^2 P_2^{LR}(B_s)
\frac{|(a_0+a_1)(a^*_0+a^*_2)|}{M_H^2} \frac{m_b m_s m_t^4}{v^6} |V_{tb} V^*_{ts}|^2\tan^2\beta,
\eea
where $a_0,a_1$ and $a_2$ are functions of the expansion parameters $\epsilon_i$ (see \cite{D'Ambrosio:2002ex} for their expression), $P_2^{LR}$ are hadronic matrix elements and are given e.g. in \cite{Buras:2001ra}. $M_H$ the mass of the heavy Higgs boson that is close to the mass of the pseudoscalar, $A$, in the alignment or decoupling limit $\cos(\alpha-\beta)=0$. An analogous expression holds for the $B_d$ system. Additional NP contributions can arise from the exchange of the light Higgs boson, $h$, but these are generically sub-dominant, as they are not enhanced by $\tan\beta$. These expressions show that larger NP effects arise in the $B_s$ system, $\Delta M_{B_s}^{\rm{NP}}\gg \Delta M_{B_d}^{\rm{NP}}\gg \Delta M_K$,  and that the NP contributions have the same dependence on the quark masses and CKM elements, as in the SM. This particular structure leads to not too strong constraints on the heavy Higgs boson masses. Even in the case of $\mathcal O(1)$ phases in the $\epsilon_i$ parameters, one finds the condition $\tan\beta (v/M_H)<$ few, leading to EW scale heavy Higgs bosons, in the case of not too large values of $\tan\beta$ \cite{Buras:2010mh}.

Similarly, Higgs exchange tree-level diagrams contribute to the rare $B_{s,d}\to\mu^+\mu^-$ decays. If we assume the decoupling (or alignment limit), $\cos(\alpha-\beta)=0$, and $m_H=m_A$, then the pseudoscalar and scalar contributions are the same and the branching ratios of the $B_{s,d}$ rare decays read \cite{Buras:2010mh}
\beq\label{eq:BmumuMFV2HDM}
\frac{{\rm{BR}}(B_{s,d}\to\mu^+\mu^-)}{{\rm{BR}}(B_{s,d}\to\mu^+\mu^-)_{\rm{SM}}}\simeq|1+R_{s,d}|^2+|R_{s,d}|^2,
\eeq
with
\beq\label{eq:RMFV2HDM}
R_{s,d}=(a_0^*+a_1^*)\frac{2\pi^2m_t^2}{Y(x_t)m_W^2}\frac{m_{B_{s,d}}^2\tan^2\beta}{(1+m_{s,d}/m_b)M_H^2},
\eeq
where we have neglected the (small) contribution of the lightest Higgs, $h$, that is not $\tan^2\beta$ enhanced. It is straightforward to demonstrate that the branching ratios predicted by this MFV 2HDM obey to the relation in (\ref{eq:ratioBmumuMFV}), modulo corrections proportional to the ratios of masses $m_{s,d}/m_b$. These corrections are, however, well below the parametric uncertainties on the SM predictions for the two branching ratios. Using Eqs. (\ref{eq:BmumuMFV2HDM}) and (\ref{eq:RMFV2HDM}), one can place constraints from the measurements of $B_s\to\mu^+\mu^-$ in the famous $m_A-\tan\beta$ plane. As shown in e.g. \cite{Altmannshofer:2012ks}, these constraints are complementary to the constraints that arise from the LHC direct searches of heavy new scalar/pseudoscalars (e.g. searches for $pp\to H,A\to\tau^+\tau^-$ \cite{Aaboud:2016cre,CMS:2016pkt}).

\subsection{Flavor breaking in the SUSY models}\label{sec:SUSY}
In spite of the (so far) LHC null-results in searching for TeV-scale SUSY, Supersymmetry remains one of the best motivated theories beyond the SM.
The particle content of the Minimal Supersymmetric Standard Model (MSSM) consists of the SM gauge and fermion fields plus a scalar partner
for each quark and lepton (squarks and sleptons) and a spin-1/2 partner for each gauge field (gauginos).
The Higgs sector has two Higgs doublets with the corresponding spin-1/2 partners (Higgsinos). Similarly to the SM (see Sec. \ref{sec:SMFlavor}), the MSSM Supersymmetry preserving Lagrangian is completely determined by symmetry principles and it has a relatively small set of free parameters. However, to make the MSSM phenomenologically viable, one also has to introduce soft SUSY breaking terms. The most general soft SUSY breaking Lagrangian that is gauge invariant and respects R-parity reads
\bea\nonumber
\mathcal L_{\rm{soft}}&=&\frac{1}{2}M_1\lambda_B\lambda_B+\frac{1}{2}M_2\lambda_W\lambda_W+\frac{1}{2}M_3\lambda_g\lambda_g-m_{H_d}^2|H_d|^2-m_{H_u}^2|H_u|^2\\
&-&\tilde m_Q^2\tilde Q_L^*\tilde Q_L-\tilde m_D^2\tilde d_R^*\tilde d_R-\tilde m_U^2\tilde u_R^*\tilde u_R-\tilde m_L^2\tilde \ell_L^*\tilde \ell_L-\tilde m_E^2\tilde e_R^*\tilde e_R\\\nonumber
&+&B\mu H_uH_d+\hat A_\ell\tilde\ell H_d\tilde e_R^*+\hat A_D\tilde qH_d\tilde d_R^*-\hat A_U\tilde qH_u\tilde u_R^*,
\eea
with $M_1,M_2,M_3$ Majorana masses for the gauginos and $m_{H_d},m_{H_u}$ soft masses for the two Higgs boson doublets. In all generality, the squark and slepton soft masses ($\tilde m_Q,\tilde m_D,\tilde m_U$,\\ $\tilde m_L,\tilde m_E$) as well as the trilinear couplings ($\hat A_\ell,\hat A_D,\hat A_U$) are $3\times 3$ matrices in flavor space and introduce an additional very large number of free parameters (33 new angles and 47 new phases, of which 2 can be rotated away by field redefinitions). These soft terms lead to gluino, Higgsino and gaugino flavor changing couplings. It has been shown that low energy flavor measurements lead to bounds on the squark masses up to $10^3$ TeV in the case of a completely generic flavor structure (see e.g \cite{Altmannshofer:2013lfa}). In other words, in the case of TeV-scale SUSY, the rich flavor structure of the MSSM generically leads to large contributions to FCNC processes in conflict with available experimental data: the so-called {\it{SUSY flavor problem}}. Several models that address this problem have been proposed in the literature: models with mechanisms of SUSY breaking with flavor universality, such
as in gauge mediation models \cite{Giudice:1998bp}, models with heavy squarks and sleptons, such as in (mini) split-SUSY \cite{ArkaniHamed:2004fb,Giudice:2004tc,ArkaniHamed:2004yi,Hall:2011jd,Arvanitaki:2012ps,ArkaniHamed:2012gw}, or models with alignment of quark with squark mass matrices \cite{Nir:1993mx}.

MFV represents an interesting alternative. The MFV hypothesis can easily be implemented in the MSSM framework. The squark mass terms and the
trilinear quark-squark-Higgs couplings can be expressed as follows
\begin{eqnarray}
{\tilde m}_{Q}^2 &=& {\tilde m}^2 \left( a_1 {1 \hspace{-.085cm}{\rm l}} 
+b_1 Y_u Y_u^\dagger +b_2 Y_d Y_d^\dagger 
+b_3 Y_d Y_d^\dagger Y_u Y_u^\dagger +\ldots
 \right)~, \nonumber  \\\nonumber
{\tilde m}_{U}^2 &=& {\tilde m}^2 \left( a_2 {1 \hspace{-.085cm}{\rm l}} 
+b_5 Y_u^\dagger Y_u +\ldots \right)~,  \\\label{eq:MFVSoft}
{\tilde m}_{D}^2 &=& {\tilde m}^2 \left( a_3 {1 \hspace{-.085cm}{\rm l}} 
+b_6 Y_d^\dagger Y_d +\ldots \right)~,
\\\nonumber
\hat A_U &=&~ \tilde A\left( a_3 {1 \hspace{-.085cm}{\rm l}} 
+b_6 Y_d Y_d^\dagger +\ldots \right) Y_u~,\\
\hat A_D &=& \tilde A\left( a_5 {1 \hspace{-.085cm}{\rm l}} 
+b_8 Y_u Y_u^\dagger +\ldots \right) Y_d~,\nonumber
\label{eq:MSSMMFV}
\end{eqnarray}
with the parameters $\tilde m$ and $\tilde A$ that set the the overall scale of the soft-breaking terms and the dimensionless coefficients $a_i$ and $b_i$ generic $\mathcal O(1)$ free complex parameters of the model. The several soft masses and trilinear terms are described by a matrix proportional to the identity plus (small) corrections, suppressed by small Yukawa couplings and CKM elements. 

The NP effects in low energy flavor observables can, therefore, be computed using the so-called mass insertion approximation \cite{Hall:1985dx}. More specifically, every observable can be expressed by an expansion in $\delta =\Delta /\tilde m^2$, with $\Delta$ the off-diagonal terms in the sfermion mass matrices (proportional to the small Yukawas in the case of MFV). Using this method, one can demonstrate that, with the flavor structure in (\ref{eq:MFVSoft}) and the corresponding one in the down sector, squark masses $\tilde m$ at around the TeV scale are still consistent with flavor constraints \cite{Altmannshofer:2007cs}. We can then conclude that, if MFV holds, the present bounds on
FCNCs do not exclude squarks in the LHC reach. LHC squark direct searches and low energy flavor observables are, therefore, two complementary probes of MFV SUSY models.

\subsection{Top and Higgs flavor violating signatures}
So far in these lectures, we have discussed low energy flavor observables that have been/will be measured by B-factories and by the LHCb. High energy flavor measurements by the ATLAS and CMS collaborations provide a complementary tool to test the underlying flavor structure of Nature. Particularly, in the last few years, a tremendous progress has been achieved in the measurement of Higgs and top flavor violating couplings. This is the topic of the last section of these lectures.

The top quark is the only quark whose Yukawa coupling to the Higgs boson is order of unity and the only one with a mass larger than the mass of the weak gauge bosons. Thanks to its heavy mass, the top mainly decays to a $W$ boson and a bottom quark, with an extremely small life time of approximately $5 \times 10^{-25}$ s. This is shorter than the hadronization time, making it impossible for the top quark to form bound states.
For these reasons the top quark plays a special role in the Standard Model and in many BSM extensions thereof. 
 An accurate knowledge of its properties can bring key information on fundamental interactions at the electroweak scale and beyond.
  So far, the flavor conserving properties of the top are known with a very good accuracy. Less is know about the flavor changing top couplings. 
 
 The flavor changing decays of the top quark are suppressed by the GIM mechanism, similarly to what happens to the other quarks. The decay of a top quark to a Z boson or a photon and an up or charm quark occurs only through higher-order diagrams. These processes should be compared to the tree-level decay  to a W boson and a bottom quark, resulting in tiny top flavor changing branching ratios in the framework of the SM. In the second column of Tab. \ref{Tab:TopSMAndBounds}, we present the SM predictions for the flavor changing branching ratios of the top. All branching ratios are below the $10^{-13}$ level! A discovery of a flavor violating top decay in the foreseeable future would, therefore, unequivocally, imply the existence of New Physics. 
 
\begin{table}[t]
\begin{center}
\renewcommand{\arraystretch}{1.2}
\begin{tabular}{||c||c |c|c ||}
\hline\hline
\rule{0pt}{1.2em}
Decay mode & SM prediction & LHC bound & Comments and References\\
\hline
BR$(t\to ch)$ & $3\times 10^{-15}$ & $4.6\times 10^{-3}$ & $h\to$ lept. \cite{CMS:2015qta}, $h\to b\bar b$ \cite{Aad:2015pja,CMS:2015qhe},  $h\to\gamma\gamma$ \cite{CMS:2015xqa} \\
\hline
BR$(t\to uh)$ & $2\times 10^{-17}$ & $4.2\times 10^{-3}$ & $h\to b\bar b$ \cite{Aad:2015pja,CMS:2015qhe}, $h\to\gamma\gamma$ \cite{CMS:2015xqa}\\
\hline
BR$(t\to cg)$ & $5\times 10^{-12}$ & $2\times 10^{-4}$ & Single top production \cite{Aad:2015gea}\\
\hline
BR$(t\to ug)$ & $4\times 10^{-14}$ & $4\times 10^{-5}$ & Single top production \cite{Aad:2015gea}\\
\hline
BR$(t\to uZ)$ & $8\times 10^{-17}$ & $1.7\times 10^{-4}$ & $Z\to\ell\ell$ \cite{Aad:2015uza,Chatrchyan:2013nwa}, $tZ$ production \cite{CMS:2016bss}\\
\hline
BR$(t\to cZ)$ & $10^{-14}$ & $2\times 10^{-4}$ & $Z\to\ell\ell$ \cite{Aad:2015uza,Chatrchyan:2013nwa}, $tZ$ production \cite{CMS:2016bss}\\
\hline
BR$(t\to u\gamma)$ & $4\times 10^{-16}$ & $1.3\times 10^{-4}$ & Single top production \cite{Khachatryan:2015att}\\
\hline
BR$(t\to c\gamma)$ & $5\times 10^{-14}$ & $1.7\times 10^{-3}$ & Single top production \cite{Khachatryan:2015att}\\
\hline\hline
\end{tabular}
\caption{\label{Tab:TopSMAndBounds}SM prediction for the several flavor changing top decay branching fractions (from \cite{AguilarSaavedra:2004wm}). Also shown the present LHC bounds, as well as a few details about the searches and the corresponding reference.}
\end{center}
\end{table}

Several searches for top flavor changing couplings have been performed at the LHC, and, so far, there is no evidence for non zero couplings. In the third column of Tab. \ref{Tab:TopSMAndBounds} we show the state of the art of the most stringent constraints on the several branching ratios. All searches have been performed using the full 8 TeV luminosity. Some searches look directly for top flavor changing decays; some other for single top production, eventually in association with a $Z$ or a photon. Projections of these constraints for the HL-LHC show that we could reach the sensitivity to flavor changing branching ratios at the level of BR$(t\to gc)\lesssim 4\times 10^{-6}$ and BR$(t\to hq)\lesssim 2\times 10^{-4}$ \cite{Agashe:2013hma}. These values are still quite larger than the corresponding SM predictions, but will be crucial for testing the prediction of Randall-Sundrum models \cite{Randall:1999ee} and of 2HDMs with a generic flavor structure, that can predict branching ratios as large as BR$(t\to gc)_{\rm{2HDM}}\sim 10^{-5}$ and BR$(t\to hq)_{\rm{2HDM}}\sim 2\times 10^{-3}$, BR$(t\to hq)_{\rm{RS}}\sim 10^{-4}$, in agreement with the present low energy flavor constraints \cite{Atwood:1996vj,Agashe:2006wa}.

As we have discussed in Sec. \ref{sec:SMFlavor}, the Higgs is intrinsically connected to the flavor puzzle, as without Yukawa interactions the SM flavor symmetry, $\mathcal G_{\rm{flavor}}$, would be un-broken. For this reason, it is of paramount importance to test the couplings of the Higgs with quarks and leptons at the LHC. By now, we know that the masses of the
third generation quarks and leptons are largely due to the 125 GeV Higgs, as indicated
by the measured values of Higgs couplings to the third
generation fermions. Little is known about the origin of the masses
of the first and second generation fermions and about flavor changing Higgs couplings.

In the SM, in spite of the very small Higgs width, flavor violating Higgs decays have a negligible branching ratio. Generically, flavor violating Yukawa couplings are well constrained by the low energy FCNC measurements \cite{Blankenburg:2012ex,Harnik:2012pb}. A notable exception are the flavor violating couplings involving a tau lepton. Models with extra sources of EWSB, can predict a sizable ($\%$ level) Higgs flavor violating decays to a tau and a lepton, while being in agreement with low energy flavor observables, as $\tau\to\mu\gamma$ \cite{Altmannshofer:2015esa}.

A few searches for Higgs flavor violating decays $h\to\tau\mu,~h\to\tau e$ have been performed by the LHC \cite{Khachatryan:2015kon,Khachatryan:2016rke,CMS:2016qvi,CMS:2015udp,Aad:2015gha}, so far not showing a convincing evidence for non-zero branching ratios (see, however, the initial small anomaly shown by the CMS collaboration in \cite{Khachatryan:2015kon}). It will be very interesting to monitor these searches in the coming years of the LHC, as they could give a complementary probe of models with sizable flavor changing Higgs couplings to leptons.

\section{Summary}\label{sec:conclusions}
An essential feature of flavor physics is its capability to probe very high scales,
beyond the kinematical reach of high energy colliders. At the same time, flavor physics can teach us about properties of TeV-scale new physics (i.e. how new particles couple to the SM degrees of freedom), offering complementarity with searches of NP at colliders.

In these lectures, I discussed the flavor structure of the SM, particularly focusing on the symmetry principles of the SM Lagrangian and on how the flavor symmetry is broken. Flavor changing neutral processes in the SM are highly suppressed, both because they arise at least at the loop-level and because of the GIM mechanism that introduces the dependence of these processes on the small CKM off-diagonal elements and on the small quark masses. 

Due to the SM suppression of FCNC processes, flavor transitions offer a unique opportunity to test the New Physics flavor structure. Generically NP models predict too large contributions to flavor transitions (the ``New Physics flavor problem") leading us to conclude that, if TeV-scale New Physics exists, it must have a highly non generic flavor structure, as for example it can obey to the Minimal Flavor Violation principle. 

Several experiments are running and will be running in the coming years (LHCb, Belle II, NA62, KOTO and many lepton flavor experiments) and many more observables will be measured precisely. Some of the golden channels for the coming years are
\begin{itemize}
\item More precise measurement of  the clean rare decays $B_s\to\mu^+\mu^-$ and $B_d\to\mu^+\mu^-$ at LHCb, ATLAS and CMS. The ratio of branching ratios will give us more insights on the validity of the MFV ansatz.
\item Additional tests of the lepton universality relations in $B$ decays at LHCb and Belle II: BR$(B\to J ee)/{\rm{BR}}(B\to J \mu\mu)$ with $J=K,K^*,X_s,K\pi,...$ . These are particularly clean tests of the SM, as the theory predictions are known to a very good precision and are not affected by hadronic uncertainties.
\item Better measurements of $B\to D\tau\nu$ and $B\to D^*\tau\nu$, to confirm or disprove the present anomaly in these decays, as observed at Belle, Babar and LHCb \cite{Lees:2012xj,Lees:2013uzd,Aaij:2015yra,Huschle:2015rga}. 
\item Brand new measurements of $B\to K^{(*)}\nu\nu$ and $K\to\pi\nu\nu$ at Belle II and KOTO, respectively. 
\item Additional searches of top and Higgs flavor violating couplings at the LHC.
\end{itemize}
These channels (and several others) will be able either to set interesting constraints on NP, or to shed light into the existence of new degrees of freedom beyond the SM.

\section*{Acknowledgements}
I would like to thank Wolfgang Altmannshofer and my student, Douglas Tuckler, for comments on the manuscript. 
I acknowledge support from the University of Cincinnati. I wish to thank the organizers of the 2015 European School of High-Energy Physics in Bansko for the invitation. I am also grateful to the students and the discussion leaders for interesting discussions.


\end{document}